\documentclass[final,5p,times,twocolumn]{elsarticle}
\usepackage{graphicx}
\usepackage{float}
\usepackage{caption}
\usepackage{subcaption}
\usepackage{tikz}
\usepackage{siunitx}
\usepackage{amsmath}
\usepackage{amssymb}
\usepackage{amsthm}
\usepackage{booktabs}
\usepackage{wrapfig}
\usepackage[colorlinks=true,linkcolor=blue,anchorcolor=blue,citecolor=blue,filecolor=blue,menucolor=blue,runcolor=blue,urlcolor=blue,breaklinks]{hyperref} 
\usepackage{svg}
\usepackage{lipsum} 
\usepackage{placeins} 
\usepackage{chemformula} 
\usepackage{tabstackengine} 
\stackMath
\usepackage{lettrine} 
\usepackage{parskip}
\usepackage{amsmath,amsfonts,amssymb} 

\usepackage{amsthm}                   
\usepackage[graphicx]{realboxes}
\usepackage{varwidth}
\usepackage[ruled,vlined]{algorithm2e}
\usepackage{algorithmic}
\usepackage{subcaption} 
\usepackage{listings}                 
\makeatletter

\begin{document} 
\begin{frontmatter}

\title{Digital Twins in Wind Energy: Emerging Technologies and Industry-Informed Future Directions}
\journal{Published in IEEE Access}

\author[NTNUIKT]{Florian Stadtmann}
\author[NTNUIKT,SINTEFD]{Adil Rasheed\corref{mycorrespondingauthor}}
\ead{adil.rasheed@ntnu.no}
\cortext[mycorrespondingauthor]{Corresponding author}
\author[NTNUIMF,SINTEFD]{Trond Kvamsdal}
\author[SINTEFD]{Kjetil Andr\'{e} Johannessen}
\author[OSU]{Omer San}
\author[SINTEFE]{Konstanze Kölle}
\author[SINTEFE]{John Olav Tande}
\author[NORCONSULT]{Idar Barstad}
\author[TE]{Alexis Benhamou}
\author[VARD]{Thomas Brathaug}
\author[DNV]{Tore Christiansen}
\author[SUSE]{Anouk-Letizia Firle}
\author[FORCE]{Alexander Fjeldly}
\author[4SUBSEA]{Lars Frøyd}
\author[COGNITE]{Alexander Gleim}
\author[EIDEL]{Alexander Høiberget}
\author[MAINSTREAM]{Catherine Meissner~}
\author[STORENORSK]{Guttorm Nygård}
\author[STATKRAFT]{Jørgen Olsen}
\author[KDI]{Håvard Paulshus}
\author[ANEO]{Tore Rasmussen}
\author[DNV]{Elling Rishoff}
\author[Equinor]{Francesco Scibilia}
\author[KM]{John Olav Skogås}

\address[NTNUIKT]{Department of Engineering Cybernetics,Norwegian University of Science and Technology, O. S. Bragstads plass 2, Trondheim, NO-7034, Norway}
\address[SINTEFD]{Mathematics and Cybernetics, Sintef Digital, Klæbuveien 153, Trondheim, NO-7031, Norway }
\address[NTNUIMF]{Department of Mathematical Sciences, Norwegian University of Science and Technology, O. S. Bragstads plass 2, Trondheim, NO-7034, Norway}
 \address[OSU]{Mechanical and Aerospace Engineering, Oklahoma State University, 201 General Academic Building, Stillwater, Oklahoma 74078 USA }
 \address[SINTEFE]{Sintef Energy Research, Sem Sælands vei 11, 7034 Trondheim, Norway}
 \address[NORCONSULT]{Norconsult, Vestfjordgaten 4, Sandvika, NO-1338, Norway}
 \address[TE]{TotalEnergies, 2 place Jean Miller, 92078 Paris La Défense, France
}
 \address[VARD]{Vard, Skansekaia 2, Ålesund, NO-6002, Norway}
 \address[DNV]{DNV, Veritasveien 1, Høvik, NO-1363, Norway}
 \address[SUSE]{Sustainable Energy, Meatjønnsvegen 74, NO-5412 Stord, Norway}
 \address[FORCE]{FORCE Technology Norway AS, Sluppenvegen 19, NO-7037,  Trondheim, Norway}
 \address[4SUBSEA]{4subsea, Hagaløkkveien 26, Asker, NO-1383, Norway}
 \address[COGNITE]{Cognite, Oksenøyveien 10, Lysaker, NO-1366, Norway}
 \address[EIDEL]{EIDEL, Smed Hagens veg 11, Eidsvoll, NO-2080, Norway}
 \address[MAINSTREAM]{Mainstream Renewable Power, }
 \address[STORENORSK]{Store Norske, Postboks 613, NO-9171 Longyearbyen, Norway}
 \address[STATKRAFT]{Statkraft, Lilleakerveien 6, 0283 Oslo, Norway}
 \address[KDI]{Kongsberg Digital, Lysaker Torg 35, Lysaker, NO-1366, Norway}
 \address[ANEO]{Aneo, Klæbuveien 118, Trondheim, NO-7031, Norway}
 \address[Equinor]{Equinor ASA Arkitekt Ebbells veg 10, Rotvoll 7005 Trondheim, Norway}
 \address[KM]{Kongsberg Maritime, Kirkegårdsveien 45, Kongsberg, NO-3616, Norway}

\begin{abstract}
This article presents a comprehensive overview of the digital twin technology and its capability levels, with a specific focus on its applications in the wind energy industry. It consolidates the definitions of digital twin and its capability levels on a scale from 0-5; 0-standalone, 1-descriptive, 2-diagnostic, 3-predictive, 4-prescriptive, 5-autonomous. It then, from an industrial perspective, identifies the current state of the art and research needs in the wind energy sector. It is concluded that the main challenges hindering the realization of highly capable digital twins fall into one of the four categories; standards-related, data-related, model-related, and industrial acceptance related. The article proposes approaches to the identified challenges from the perspective of research institutes and offers a set of recommendations for various stakeholders to facilitate the acceptance of the technology. The contribution of this article lies in its synthesis of the current state of knowledge and its identification of future research needs and challenges from an industry perspective, ultimately providing a roadmap for future research and development in the field of digital twin and its applications in the wind energy industry.
\end{abstract}

\end{frontmatter}


\section{Introduction} 
\label{sec:introduction}
Wind energy is expected to play an important role in limiting global warming to the recommended preindustrial levels~\cite{IPCC}. In terms of greenhouse gas emissions, wind electricity can compete well with low-emission hydro- and photovoltaic electricity, and offshore wind energy can outperform both in the use of land~\cite{Gibon2022cni}. Wind energy has been one of the fastest-growing energy sources globally, with a 53 percent year-on-year increase in 2020~\cite{Lee2021gwr}. The EU's vision aims at a climate-neutral EU by 2050~\cite{2018cft}. According to~\cite{2018ida}, 51-56\% of the power production is planned to come from the wind in 2050 and 26\% in 2030. Following~\cite{Freeman2019oeo}, up to 30\% of the European electricity demand is planned to be met by offshore wind in 2050. This is estimated to equal 450~GW. Worldwide, the wind energy electricity generation capacity is estimated to grow by a factor of 8 until 2050, to 5~TW~\cite{Murray2019diw}. The technical potential of offshore wind in 2019 was already at 48~TW, which was 18 times more than the global demand for electricity at that time~\cite{IEA2019owo}.

Reducing both the cost and environmental footprint of wind energy is crucial to realize a zero-emission wind-powered future~\cite{2018ida}. However, the cost reduction of wind energy in the last decades happened together with a significant increase in turbine size, but without new innovations, up-sizing might not yield further benefits~\cite{Islam2013par,Ashuri2016mdo}. With cheaper sensors and computational resources, the benefits of gathering, recording, and analyzing data are growing as well. Through digitalization, the efficiency of wind farms during the whole life-cycle is increased, from design~\cite{Ashuri2016mdo} and siting~\cite{Velo2014wse}, over construction~\cite{Asgarpour2016ati} and operation~\cite{Padullaparthi2022ffl}, to maintenance~\cite{Nichenametla2017olc} and decommissioning~\cite{Irawan2019aom}. If we take into account the current fleet of wind turbines throughout the world, a 1\% increase in overall energy would result in more than 30 terawatt hours of additional electricity each year, which is roughly similar to adding an additional 3,600 wind turbines for free~\cite{Howland2022cwf}. This would result in an additional \$1 billion in revenue annually for the owners of wind farms.  This increases sustainability not only by reducing the levelized cost of electricity (LCOE), thereby increasing the competitiveness of wind energy with fossil energy, but it also reduces the number of resources needed per energy produced, thus increasing the sustainability of wind energy itself.
Simulations and recorded data from operational wind farms and prototypes aid in optimizing the wind turbine design for future turbine generations. Simulations and wind measurements allow identifying the optimal farm- and turbine-site~\cite{Velo2014wse}, as well as the optimal turbine configuration for each site~\cite{2015htd}. Digitalization also helps in planning and optimizing production steps for smooth assembly, transportation, and commissioning~\cite{Asgarpour2016ati}. Two of the most requested digital solutions are condition-based and predictive maintenance. About 30-34\% of the levelized cost of electricity in wind power is estimated to stem from Operation and Maintenance (O\&M), where catastrophic O\&M events are not even included yet~\cite{Stehly20202co}. The O\&M cost can increase by up to~95\%~\cite{Walford2006wtr} of the investment cost, which presents a high risk for operators. In addition, offshore wind farms are often located in remote and hazardous locations that are difficult to access. Condition-based maintenance reduces the number of maintenance jobs on site. Predictive maintenance goes a long way in eliminating unexpected downtime and preventing catastrophic failures by taking action before small faults can get out of hand~\cite{Nichenametla2017olc}.

Weather forecasts are based on digital tools and are used, for example, for power prediction~\cite{Chen2014wpf}. 
Manual and automated control is relevant not only for optimizing revenue but also for increasing the wind-farm lifetime~\cite{Padullaparthi2022ffl}. Condition monitoring could also help during decommissioning. The recorded data can help in identifying the wear of each component of the wind farm and thereby optimizing the recycling process. Furthermore, the decommissioning scheduling happens digitally~\cite{Irawan2019aom}. 

Digital Twins (DT) combine all the previously mentioned trends and more into a single framework at scale \cite{Niederer2021sdt,Kapteyn2021apg}.
The DT accompanies the physical asset through its entire life cycle, from design to decommissioning, and can be used even afterward for recycling and the design of future assets. The combination of big data and physically accurate simulations into a real-time updated virtual system allows applying all previously mentioned techniques with maximum information through large available data sets and in a single interface with optimal visualization.
Therefore, such a holistic DT is at the core of digitalization and builds the keystone for a successful digital future. One of the challenges with digital solutions is that they are fragmented \cite{Blair2021dto}. DT, which has become very effective in other fields like meteorology and the manufacturing industry, seems to address many of the challenges faced during digitalization in the context of wind energy. The term DT has been around for two decades~, and the concept started out from product lifecycle management \cite{Grieves2017dtm, Grieves2006plm}. A decade ago, it was first applied to high value assets such as aerospace vehicles, \cite{Glaessgen2012tdt}, \cite{ Shafto2010dms, Rosen2015ati}. During recent years DTs have been utilized by various industries such as manufacturing, healthcare, aviation, automotive transportation, infrastructure planning, and energy production including wind energy (compare Table 1 in \cite{Chen2020dti} and \cite{GEPowerDigitalSolutions2016gdt}), but not to its full capacity, and many challenges remain \cite{Tao2019mmd}. Especially in the context of wind power, DTs are greatly under-utilized~\cite{Ciuriuc2022dtf}. 

In this paper, we analyze the industry's perspective and state-of-the-art of DT. We perform a survey with~15 companies operational in the Norwegian and international wind power sectors and summarize each individual answer into a short paragraph. On the basis of the survey, we investigate the benefits the wind industry is expecting from DT technology, as well as the challenges that are encountered and anticipated while developing and applying DT technology. We build a definition and taxonomy for DT based on both industrial and academic needs. To this end, we present a taxonomy based on the capability of DT, which is inspired by the value generated from a DT at each milestone during the technology development. Furthermore, we propose solutions for the challenges raised in the industry survey. Topics include data generation, gathering, and sharing, visualization, physics-based, data-driven, and hybrid-modeling, forecasting, \textit{what-if ?}-scenario analysis, manual and automated control, and industrial acceptance.

More concretely, the current article attempts to:
\begin{itemize}
    \item Consolidate the definition of DT and its capability levels from an industrial perspective in the context of wind energy.
    \item Identify the current state of the art in the industries active in the wind energy business and research.
    \item Identify the research needs and challenges that should be prioritized from an industry perspective.
    \item Propose approaches to each of the identified challenges from the perspective of research institutes. 
    \item Define a set of recommendations for the diverse class of stakeholders to facilitate the acceptance of the technology. 
\end{itemize}
To the best of our knowledge, this is the first work in which multiple industry players active in digital-twin-related activities have been brought together to provide concrete insight into some of the most pressing challenges and their potential solutions. 

In Section~\ref{sec:commonlyusedterminologies}, we present a definition and taxonomy for DT that is used in this paper and that all authors, coauthors, and industrial partners of this paper have agreed upon. Section~\ref{sec:survey} presents the results of a survey with 15 companies that work in the Norwegian and international wind sectors. Here, we analyze the industry's perspective on which values are most desired and which challenges have to be addressed before establishing a DT. In section~\ref{sec:sota}, we give an overview of state-of-the-art technologies and trends relevant to realize a DT that addresses all industrial needs. Topics include data generation, gathering, and sharing, visualization, fast physically accurate modeling, and control. Section~\ref{sec:sota} gives recommendations to all stakeholders on how to facilitate DT development and the acceptance of the technology. Finally, in Section~\ref{sec:conclusion} we conclude the article with some recommendations for moving forward.

\section{Commonly used terminologies}
\label{sec:commonlyusedterminologies}
Since this work involves a wide spectrum of industry and research partners as well as the targeted audience, we first present a brief description of the terms that have been used in the current work. 

\subsection{Digital twin, digital sibling, and digital thread}
        \begin{figure*}[!htb]
        \begin{subfigure}{\linewidth}
	\centering
	\includegraphics[width=\textwidth]{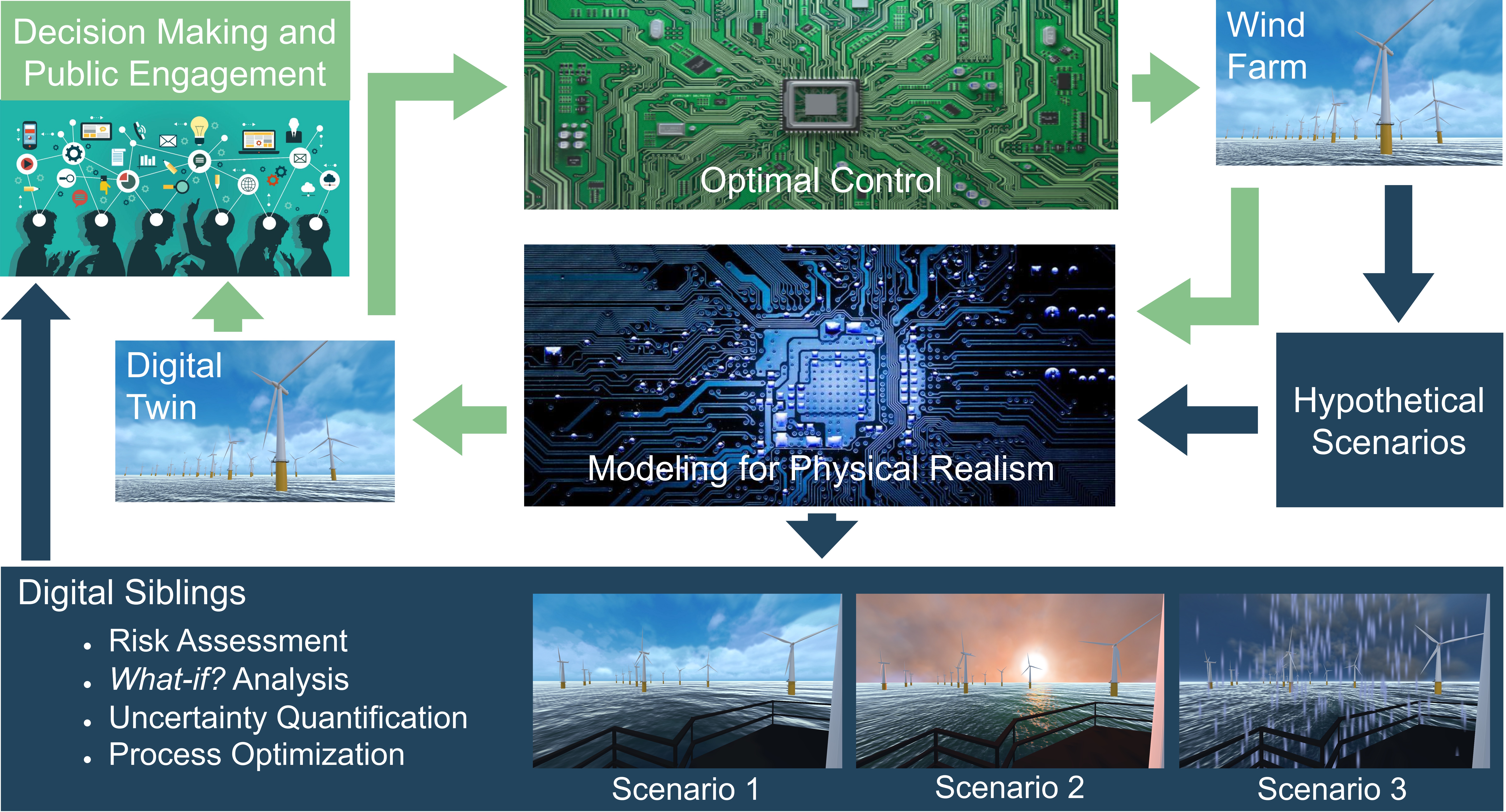}
	\caption{Overview of the Digital Twin concept applied to wind energy (adapted from~\cite{Rasheed2020dtv}).} 
	\label{fig:dt} \vspace{0.5cm}
        \end{subfigure}
        \begin{subfigure}{\linewidth}
	    \centering
	\includegraphics[width=\textwidth]{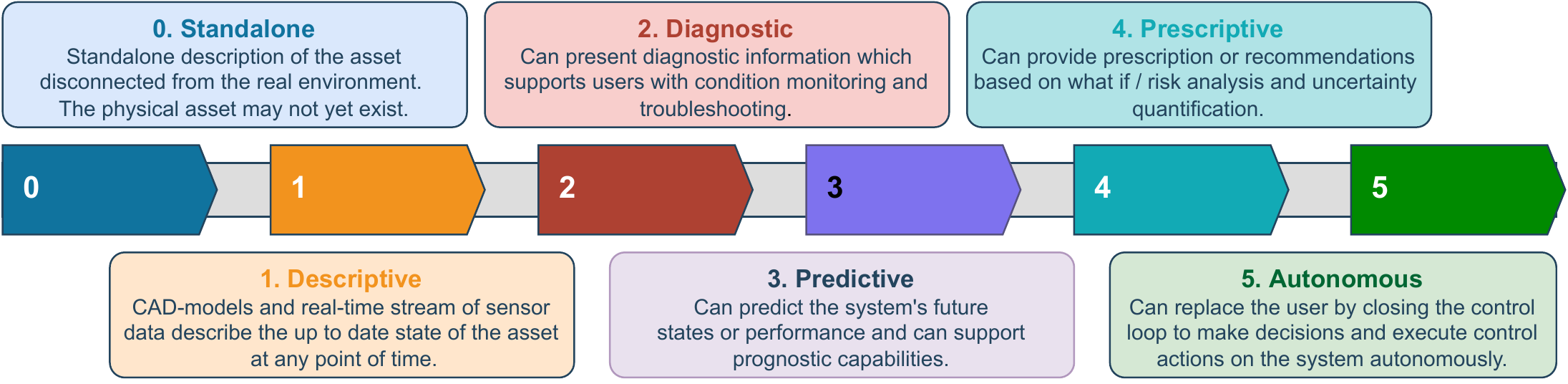}
	\caption{The capability levels of Digital Twins on a scale from 0 to 5~\cite{San2021haa}.} 
	\label{fig:capabilitylevels}
        \end{subfigure}
        \caption{Digital twin and its capability levels}
        \label{fig:DTCL}
        \end{figure*}
        
The number of definitions for the term DT is vast and varies in length. Here, we start by quoting the most popular definitions and make an attempt to redefine a long and a short concise version of the definitions.  

\begin{itemize}
    \item \textbf{Gartner} \textit{A DT is a digital representation of a real-world entity or system. The implementation of a DT is an encapsulated software object or model that mirrors a unique physical object, process, organization, person, or other abstraction. Data from multiple digital twins can be aggregated for a composite view across a number of real-world entities, such as a power plant or a city, and their related processes.} \cite{Gartnerdod}
    
    \item \textbf{NVIDIA} \textit{A digital twin is a virtual representation synchronized with physical things, people, or processes.} \cite{Martin2021wia}
    
    \item \textbf{IBM} \textit{A digital twin is a virtual representation of an object or system that spans its lifecycle, is updated from real-time data, and uses simulation, machine learning, and reasoning to help decision-making.} \cite{IMB2020wia}
    
    \item \textbf{DNV} \textit{A digital twin is a virtual representation of a system or asset, that calculates system states and makes system information available, through integrated models and data, with the purpose of providing decision support, over its life cycle.} \cite{DNVGL2020dra}
    
    \item \textbf{GE Digital} \textit{Digital Twin is most commonly defined as a software representation of a physical asset, system, or process designed to detect, prevent, predict, and optimize through real-time analytics to deliver business value.} \cite{Parriswia}
    
    \item \textbf{Siemens} \textit{A digital twin is a virtual representation of a physical product or process, used to understand and predict the physical counterpart’s performance characteristics. Digital twins are used throughout the product lifecycle to simulate, predict, and optimize the product and production system before investing in physical prototypes and assets.} \cite{Siemensdts}
    
    \item \textbf{Oracle} \textit{A digital twin is the digital proxy of a physical asset or device.} \cite{Oracleati} Alternatively, \textit{a digital twin is a digital representation of a physical asset that’s updated with operational data and is created for one asset or a fleet of assets to help in maximizing their performance. A digital twin can be built by combining real-time operational data with a physics-based model of a system or by using historical data to algorithmically determine the system’s expected behavior. Digital twins can be used for various purposes: They can provide more virtual sensor information to supplement the measured signals, help determine anomalous behavior, provide corrective actions when such behaviors occur, and even give insights to prevent anomalies from occurring in the first place. A digital twin can be created for a specific business objective across a fleet of assets, such as predictive maintenance, or for a specific piece of equipment, such as a gearbox within a larger wind turbine.}  \cite{Hutsell2022oci}
   
    \item \textbf{Microsoft} \textit{A digital twin is an exact replica of an object in the physical world that can be studied and changed to help improve the real-life version.} \cite{Microsoft2022srm}
   
    \item \textbf{Digital Twin Consortium} \textit{A digital twin is a virtual representation of real-world entities and processes, synchronized at a specified frequency and fidelity. Digital twin systems transform business by accelerating holistic understanding, optimal decision-making, and effective action. Digital twins use real-time and historical data to represent the past and present and simulate predicted futures. Digital twins are motivated by outcomes, tailored to use cases, powered by integration, built on data, guided by domain knowledge, and implemented in IT/OT systems.} \cite{Olcott2020dtc}
    
    \item \textbf{Trauer et al.} \textit{A Digital Twin is a virtual dynamic representation of a physical system, which is connected to it over the entire lifecycle for bidirectional data exchange.} \cite{Trauer2020wia}

    \item  \textbf{Grieves, Vickers} \textit{The Digital Twin is a set of virtual information constructs that fully describes a potential or actual physical manufactured product from the micro atomic level to the macro geometrical level. At its optimum, any information that could be obtained from inspecting a physical manufactured product can be obtained from its Digital Twin. Digital Twins are of two types: Digital Twin Prototype (DTP) and Digital Twin Instance (DTI). DT’s are operated on in a Digital Twin Environment (DTE).} \cite{Grieves2017dtm}

    \item \textbf{Industrial Digital Twin Association} \textit{Digital Twin: Digital representation, sufficient to meet the requirements of a set of use cases. Note: In this context, the entity in the definition of digital representation is typically an asset.} \cite{dti}

\end{itemize}
More definitions can be found, for example, in \cite{Negri2017aro}. Here, we adopt our own original definitions from \cite{Rasheed2020dtv}. \vspace{4pt}
\newline
\fbox{
	\parbox{0.95\linewidth}{
	A {\bf digital twin} is defined as a virtual representation of a physical asset or a process enabled through data and simulators for real-time prediction, optimization, monitoring, control, and informed decision-making.}
}
\vspace{2pt}

The concept can be described using Figure~\ref{fig:dt}. On the top right side of the figure, we have the physical asset /process we want to build a DT of. The asset is equipped with a diverse class of sensors that provide big data in real-time. These data have a very coarse spatio-temporal resolution and do not describe the future state of the asset. Therefore, to complement the measurement data, models are utilized to bring physical realism to the digital representation of the asset. Provided that the same information can be gained from the DT as it can be from the physical asset, it can be utilized for more informed decision-making and optimal control of the asset. All the green arrows in the figure represent real-time data exchange and analysis. However, one might be interested in risk assessment, \textit{what-if ?} analysis, uncertainty quantification, and process optimization. These can be realized by running the DT in an offline setting for scenario analysis. The concept is then known as a digital sibling. The box and the blue arrows represent the digital sibling. Additionally, the DT predictions can be archived during the lifetime of the asset and can be used for designing the next generation of assets, in which case the concept is referred to as digital threads. 
    
\subsection{Capability of digital twins}
As mentioned earlier, the concept of DTs has been around for a few decades. However, the usage of the term can be misleading. Some focus on the visual 3D component, others on the data stream between a physical asset and its digital counterpart, and others on modeling and analysis techniques. In~\cite{Solman2022dta} DTs are described as "boundary objects," entailing that the term DT has a different meaning for different stakeholders. Commonly used scales for technological progress, such as the widely known Technology Readiness Level scale (see \ref{ssec:Industrialacceptance}), are not sufficient to classify DTs. While such scales describe the technology's maturity, they do not address the features included in the technology. To this end, attempts have been made to find a scale for categorizing DTs. Grieves \cite{Grieves2017dtm} separates between Digital Twin Prototype, and Digital Twin Instance, where the former describes a product and the latter an instance of a product. In~\cite{Tygesen2018ttd} they use the term ``True" DT and introduce sub-stages for model implementation. An 8-dimensional taxonomy based on 233 DT-related publications is presented in~\cite{vanderValk2020ato}. In contrast, a distinction based on the level of automated data exchange between physical and digital objects is made in~\cite{Kritzinger2018dti}. There, a digital object without automated data exchange is called a {\em Digital Model\/}, and a digital object that only receives data is called a {\em Digital Shadow\/}. Thus, when applying the taxonomies to the state-of-the-art, it becomes evident that taxonomies are often either only focusing on a single aspect like data connection, or introducing a new dimension for each additional feature. However, these features are often dependent on each other: A data connection between an asset and its digital counterpart can only exist if the asset exists, and a bidirectional data connection only makes sense if the DT has feedback/control to provide to the asset. Therefore, we do not use any of the above classification attempts. Instead, we expand on the capability level scale as used in~\cite{Elfarri2023aid,Sundby2021gcd, DNVGL2020dra}. A one-dimensional capability-based scale brings several advantages. It uses milestones, allowing for a clear distinction between levels based on whether the milestone is reached or not and providing an improvement over empirical classifications. Each level expands on the capability of lower levels by adding new features. The levels are structured so that with each level new values can be unlocked. The scale is particularly helpful during the development stage of DTs, as it presents a progression plan and can be used simultaneously to categorize the state-of-the-art. The exact requirements for data streaming, visualization, modeling, analysis, and control depend on the DT's capability level. The scale ranges from 0 to 5 (0-standalone, 1-descriptive, 2-diagnostic, 3-predictive, 4-prescriptive, 5-autonomous) (see Figure~\ref{fig:capabilitylevels}) and is elaborated in the following sections. 
 
\paragraph{Level 0: Standalone}
A standalone DT is defined as a DT even before the physical asset comes into existence. The value of a standalone DT, in addition to being used for design purposes, is that it can be used for a preliminary cost-benefit analysis of the asset before it is built.

In the concrete case of wind farms, a standalone DT can be used for wind farm siting, wind turbine micro-siting, wind availability studies, or long-term climate studies. Another application is explored in a project by GE, where turbine components for each turbine are configured based on its site~\cite{2015htd}.

\paragraph{Level 1: Descriptive}
When geometric computer-aided design (CAD) models are in place and a live sensor data stream is established, it can be referred to as the descriptive DT, which can provide insight into the inner workings of the asset at the required granularity. Additionally, it provides a powerful interface. Numbers can be shown where relevant, critical components can be highlighted through color, rendering priority enables X-ray vision, and heat-map overlays allow visualization of parameters like stress, temperature, or component wear. Sensor data is typically only available at specific positions and times. Physically accurate models can help to interpolate data to areas of interest. Some physical assets or components thereof are challenging to access. The physical asset might be too expensive, dangerous, or inaccessible for humans. A descriptive DT mirrors the physical asset's current state and can be easily explored remotely.

A descriptive wind farm DT can be used to remotely monitor the wind farm, which is especially beneficial for wind farms in remote locations such as offshore wind farms and farms in arctic regions. In contrast to a common SCADA system, it can provide a 3D visualization of all relevant data collected from the wind farm, allowing experts to interpret the data faster and enabling non-technical staff to understand it.

\paragraph{Level 2: Diagnostic}
At a capability level of 2, data analysis tools are applied to the data for sanity-checks of sensors and data, condition monitoring and fault diagnosis. The DT is referred to as the diagnostic DT. Based on the current state of the DT, diagnostic tools such as vibration detection can be employed to detect irregularities in the DT even before they cause failures. Experts can then use the information provided by the diagnostic DT to make adjustments before minor faults result in more significant consequences.

The diagnostic DT is especially interesting for high-value assets such as wind farms, as it allows the detection of irregularities in the turbine before the irregularities can cause faults or unexpected downtimes of the turbine. Additionally, it reduces the need for on-site inspections, especially in remote locations.

\paragraph{Level 3: Predictive}
It should be noted that standalone-, descriptive- and diagnostic-DTs do not give any insight into the future. Predictive DT, as the name suggests, starts exploiting models to project the current and past states into the future. The prediction is continuously updated based on the real-time data stream from the physical asset. With the constant update of the asset state, there is no risk of diverging too far from the physical asset over time. 

The predictive DT has several applications in the context of wind farms. Short-term wind, weather, and power forecasts allow for the optimization of turbine settings. Mid-term forecasts are required for energy trading on the day-ahead energy market. Long-term forecasts aid by estimating revenue or even predicting the impact of climate change. Furthermore, the predictive DT can be used to forecast component wear and estimate the remaining useful lifetime of components throughout the wind farm, thus alleviating unexpected downtimes. Combining weather forecasts and predictive maintenance allows scheduling inspections and maintenance when there is no wind, thereby further reducing the effective downtime of the wind turbines.

\paragraph{Level 4: Prescriptive}
For optimal control of the asset, prescriptive DTs come in handy as they can make recommendations based on \textit{what-if ?} ~/ risk assessment and uncertainty quantification. This aspect is highly desirable for decision support systems, providing recommendations to experts who then decide how to act upon them.

In the context of wind farms, the prescriptive DT can improve on the diagnostic and predictive features not only to provide data with which good maintenance schedules can be developed, but directly explore \textit{what-if ?} scenarios to provide optimal maintenance schedules. Furthermore, it includes uncertainty estimates to indicate how reliable the recommendations are. Additionally, the prescriptive DT can provide recommendations for turbine settings at the farm level to prolong component and turbine lifetime.

\paragraph{Level 5: Autonomous}
Finally, the DT and the digital asset start bidirectional communication where the physical asset updates its DT in real-time, and in return, the DT controls the asset to push it towards an optimal set point. Decisions can be made on much shorter timescales than with human involvement. This autonomous DT represents the fifth level. A high level of maturity is needed in order to use autonomous DT in critical components or systems like e.g., autonomous vehicles.

The autonomous wind farm DT allows for closing the loop in the wind farm operation. It can continuously adjust the turbine settings at the farm level to optimize turbine efficiency while taking into account current and future weather, wind speed, wind direction, electricity prices, wake effects, component wear, and the remaining useful lifetime of components. With sufficient maturity, it can furthermore schedule maintenance autonomously. By exploring the use of autonomous drones and underwater vehicles, autonomous DT can eventually repair minor damage on its own.

\subsection{Informed decision making and public opinion}
Decision-making is the act or process of deciding something, especially with a group of people. A decision is an informed decision when the decision takes into account knowledge about the potential consequences of the decision and their probabilities. While in theory all decisions should be informed decisions, there are many situations where a lack of information, time, or expertise impedes informed decision-making. This poses serious challenges in industrial and political contexts, where decisions can have far-reaching consequences.

Public opinion is understood as an aggregate of individual attitudes or beliefs about a particular topic or issue held by a significant proportion of the total population. Public opinion has a strong influence on decisions in society on all scales, from the behavior of individual citizens to global politics. However, public opinions can be based on prejudice, misconceptions, and false facts, as often only a small fraction of the population possesses expertise in a topic. An informed public opinion is a public opinion that takes into account a significant amount of correct information and no incorrect information that could change the opinion. In order for public opinions to be informed, correct information needs to be easily available for a sufficiently large fraction of the population, and false information has to be identifiable as such. In many cases, this requires technical information to be understandable for individuals without a technical background in the corresponding topic.  

\section{Industrial perspective}
\label{sec:survey}
To understand the industry needs, research challenges, and potential value that can be generated from DT, industry partners of the FME NorthWind research center were familiarized with the terminologies discussed in Section \ref{sec:commonlyusedterminologies}. NorthWind - Norwegian Research Centre on Wind Energy – is a strategic precompetitive research cooperation co-financed by the Research Council of Norway, industry, and research partners. The primary objective of NorthWind is to bring forward outstanding research and innovation to reduce the cost of wind power and facilitate its sustainable development. This will grow exports and create new jobs. 

\begin{figure}
	\centering
	\includegraphics[width=\linewidth]{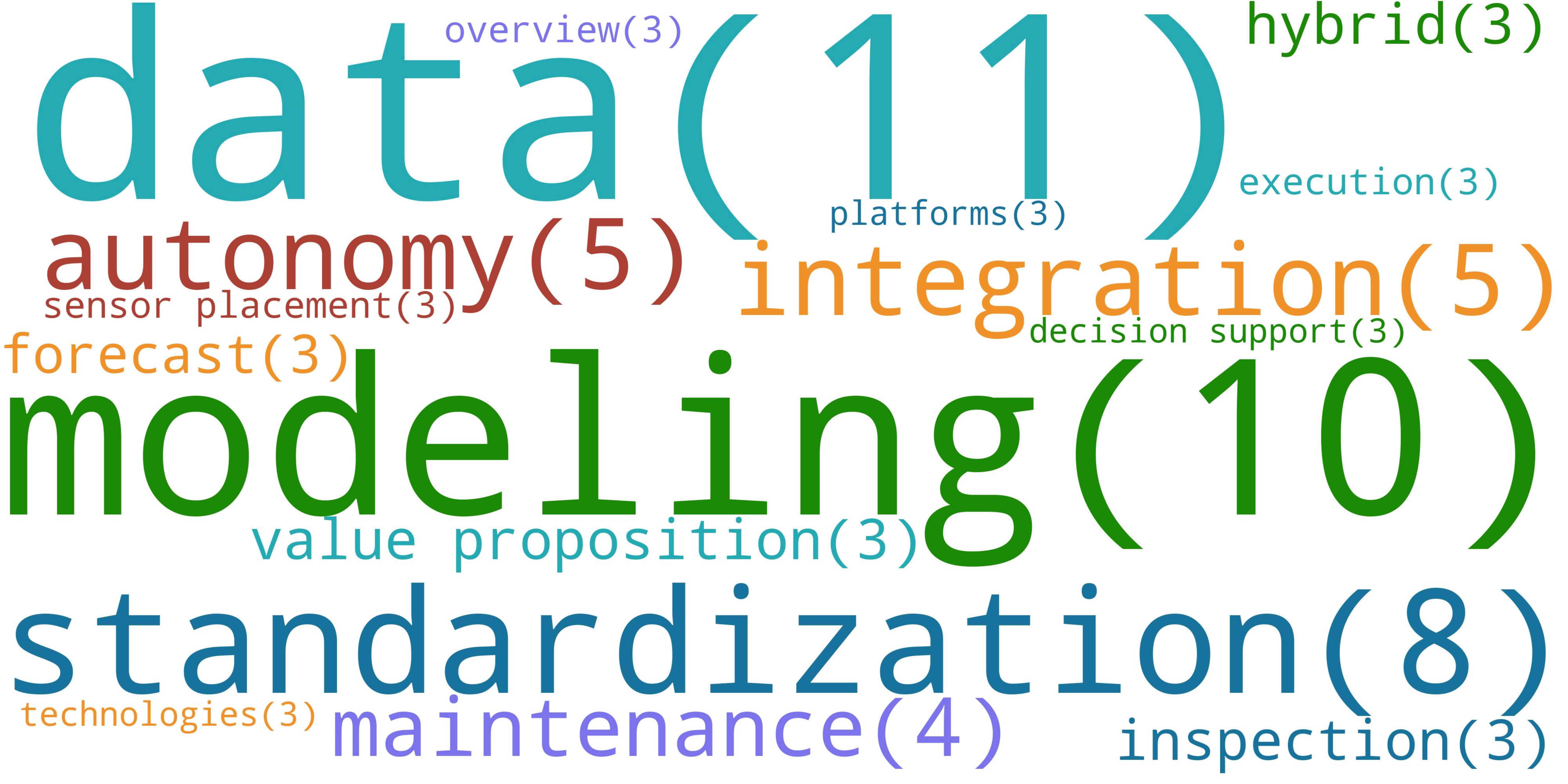}
	\caption{Recommendations from industry for academic research to focus on.
    Focus on propositions on value generation from data through combination with models has been the most frequent recommendation. The application of such hybrid models spans forecasts, \textit{what-if ?} scenario analysis, decision support, inspection and maintenance applications, and autonomous systems.
    It was mentioned that industrial partners should be included to ensure validation of the results and that digital twins should be built on real cases.
    Academic research should furthermore be involved in the standardization of data and implementation.
    Finally, overviews of applicable technologies have been requested.
 } 
	\label{fig:industriesperspective_academic}
    \end{figure}

The survey to understand challenges (Figure~\ref{fig:industriesperspective_challenges}), state-of-the-art, and recommendations (Figure~\ref{fig:industriesperspective_academic}) was split into four stages. In the first stage, 15 industry partners anonymously answered a series of questions that can be found in the appendix. In stage two, the answers were summarized and reviewed internally by industry partners. In stage three, a paragraph for each industry partner was compiled together with a representative from the industry partner. The resulting paragraphs are presented in Section \ref{ssec:industry_feedback}. In the final stage, the commonly reported challenges that must be addressed for digital twins to be applied commercially are reported in Section~\ref{ssec:common_challenges}. 

\subsection{Individual Feedback}
\label{ssec:industry_feedback}
    \subsubsection{4subsea} 
    4subsea offers decision support services to energy providers based on data analytics and digital services. 4subsea's short-term goals for DT focus on lifetime estimate and extension of all components of the wind farm. Long-term goals shift to operational decision support for optimized uptime and productivity. 4subsea argues that modeling only parts of the turbine is usually not sufficient, as the global loads are of main interest. Therefore, it might be sufficient to start with a complete but less detailed DT. Enhancing the asset with more sensors is required, but existing sensors should be utilized as much as possible. 4subsea already extracts value from a level 0 DT (standalone) to level 4 (predictive). 4subsea has a one-degree-of-freedom DT already running to capture the tower base strain of an onshore wind farm. 4subsea is able to perform lifetime estimates at different locations of the tower, even at places without sensors. However, it is agreed that a similar DT for an offshore wind farm would be much more complex. For offshore oil and gas (O\&G) 4subsea has operated DT models up to Level 4 for several years to make short-term predictions and decision support for wellhead fatigue during drilling and workover operations using subsea motion sensors on the blowout preventer. The system is fully commercialized under the name SWIM. 4subsea also has DT of 3 jacket platforms in the North Sea with a focus on integrity monitoring and anomaly detection and DT of a floating production storage and offloading in Brazil with a focus on riser fatigue integrity. 4subsea implements the DT by creating a structural model in OrcaFlex and running it in the platform 4insight.io on Azure but comments that a polished interface between OrcaFlex and insight.io is lacking right now. They also utilize machine learning models to make future predictions. The reason 4subsea's DT are not more complex yet are a lack of time and effort in system identification and model tuning of the DT based on measurements.

    \subsubsection{Aneo} 
    Aneo is a Norwegian energy company within upstream renewable energy production and downstream green energy consumption. Aneo will use the new opportunities DTs offer to reduce the cost of renewable power production within wind, hydro, hydrogen, biogas, and solar. The most significant benefits of DTs for wind power production are expected to come from the reduction of downtime and increase in the technical availability of the assets. Additionally, DTs will give insight into the decisions of prolonging the lifetime of wind farms. Aneo generates value from diagnostic (level 2) DTs onward. Nonetheless, they also focus on implementing 3D visualization to simplify the communication between analysts and operators and to improve the planning process of maintenance. In addition to the development of DTs, Aneo puts a lot of effort into the robustness of data management and bundling as these are essentials to achieve reliable results. Aneo has developed DTs with predictive capabilities for the main components in the drive train of wind turbines. Aneo is interested in autonomous DTs, but they still believe in some human input before a final maintenance decision is taken. They assume that DTs will aid the workforce by reducing the time troubleshooting takes inside the wind turbine. From their perspective, the biggest challenge is to build a system that provides confidence-inspiring results in daily operations. Aneo recommends academic research to focus on integrating \textit{what-if ?} scenarios in predictive analysis. One key factor for success is to extend research on real data and in close collaboration with the industry to verify the results.
    
    \subsubsection{Cognite} 
    Cognite is a global industrial SaaS company driving the full-scale digital transformation of asset-heavy industries by providing simple access to trustworthy and contextualized data. Cognite’s flagship product, Cognite Data Fusion, leverages DTs to empower anyone to use data to solve industrial problems with speed and ease by combining data in simulations and predictions for improved decision-making. The most important features of a DT are \textit{what-if ?} scenario analysis and autonomous decision making. Modeling only parts of the asset is disfavoured, as the focus should be on a holistic assessment. Cognite expects that additional sensors, e.g. strain gauges for analysis of structural loadings, will have to be installed on the asset in addition to the already implemented sensors. According to Cognite, they have an operational predictive DT (level 3). The DT generates value from level 0 (standalone) onward. Cognite’s current challenge is in developing and improving the \textit{what-if ?} scenario analysis. Furthermore, there is interest in the data querying schemes of industrial partners. Convincing original equipment manufacturers to override the control parameters through a DT is what Cognite anticipates as the biggest challenge for the future. 

        \subsubsection{DNV Maritime} 
    DNV Maritime is part of DNV, a classification society.
    DNV Maritime's main interest in DT is in the class status of ships but they want to digitalize oil, gas, and offshore wind as well. The most important aspects of the DT are 3D visualization and condition monitoring to improve class recommendations for customers and obtain a better holistic overview of the asset status. DNV Maritime can already extract value from a standalone DT, and in some cases, it is sufficient to model only parts of the asset. It is expected that additional hardware is required in the future. DNV Maritime is already using a DT as part of the class production for safety at sea. About 5-10 full-time equivalents are working on this DT. The DT includes simulations and \textit{what-if ?} scenarios but has no real-time capabilities yet. A real-time initiative is in development. DNV Maritime expects the biggest challenge to be end-to-end value chain support, and the existing DT is limited by a lack of business and operating models. The most important standard is considered to be ISO 15946 and a lot of value is expected from other companies' DT if assuming good standardization. DNV Maritime's recommendation for academic research is to establish an overview of applicable technologies, standards, and emerging execution platforms in relevant industries (especially offshore wind). For the future DNV Maritime predicts the workforce to focus less on modeling and more on monitoring and learning.
    
    \subsubsection{DNV Energy Systems} 
    DNV Energy Systems is part of DNV, an energy consulting company.
    DNV Energy's goal for DT is to digitalize wind turbines for condition monitoring and future prediction. Value is expected to be generated with a DT from level 3 (predictive) onward. DNV Energy uses DNV's Forecaster platform to predict energy production based on historic and live data for wind and solar power as well as power demand. The biggest challenge is expected to be the development and the capability to respond to customer requests during the initial phase. DNV Energy recommends putting less focus on the algorithms but instead focusing on the implementation, including software development, and intermittently and inconsistent data, which requires significant investment.
    
    \subsubsection{EIDEL} 
    EIDEL is a designer and supplier of rugged electronics and systems with requirements to operate under harsh environments in remote and inaccessible areas. EIDEL’s areas of expertise are remote sensing, telemetry/data acquisition, remote control, and secure communication (including encryption). Their customers are within the defense and space sector, in addition to the marine and offshore industries.
    EIDEL’s main motivation and interest in DT is to adapt its existing Data Acquisition System (designed for the defense sector), to meet the needs and requirements for remote sensing, digitizing, and data acquisition in offshore wind assets and infrastructure. EIDEL argues that accurate time between different measurement/data points is critical for event correlation and is generally one of the most important factors and data quality dimensions in the Industrial Internet of Things (IIoT). In addition, monitoring should be done not only on the asset in question and its critical parts (including structures, gears, blades, and mooring lines), but also on external environments. This is to better understand the cause and effect, both in real-time and over time, especially for predictive maintenance (a topic which is supported by \cite{Haarman2018pm4}). EIDEL's goal is to provide higher data quality and provide new insights for the optimization of asset design, optimized operational settings, better profitability, prevented downtime, and reduced maintenance trips. The descriptive DT (level 1) is the minimum level to generate value with the data acquisition system for EIDEL, but collaboration partners, lab facilities for equipment tests, and a list of requirements to meet are currently missing. EIDEL does not aim at building a complete DT alone, but is committed to providing the hardware and software for data acquisition so that partners can build their DT upon it.
    Tasks that need to be addressed are collaboration, specifically, forming research and development partnerships, and aligning on mutual goals. The requirements for offshore environments and condition monitoring have to be defined. Data acquisition-specific challenges include producing and configuring hardware and software, and determining requirements for adaption to existing systems. Furthermore, EIDEL comments that the integration of new systems within existing systems might collide with proprietary information.

    \subsubsection{FORCE Technology Norway} 
    FORCE Technology Norway is a consulting and service company.
    FORCE Technology Norway expects many advantages through DT technology, including increased precision in calculations, reduced inspection and maintenance costs, optimized service intervals, and a single, complete, and digital hierarchical asset structure that collects all information. The key features of all DT levels are considered equally important. It is expected to extract value from a descriptive DT (level 1) onward.
    FORCE Technology Norway is able to run manual predictions and \textit{what-if ?} scenarios for several assets in Oil and Gas, but are not based on a real-time data stream yet. Microsoft Azure and Google Cloud are used for data transfer and Ansys, Orcaflex, Sesam, and in-house Python code for analysis. The accuracy and reliability of automated finite element approaches and autonomous data processing are seen as the biggest challenges in the development of DT. According to FORCE Technology, academic research should focus on automated hybrid modeling and analysis. 

    \subsubsection{Kongsberg Digital}  
    Kongsberg Digital is part of the Kongsberg Group and provides software and digital solutions. Kongsberg Digital builds a DT for marine oil and gas on the Kognitwin platform with capability level 3 (predictive), which is not limited to any cloud vendor and is implemented for multiple plants.
    Kongsberg Digital wants to use that knowledge to build a DT of a wind farm, including the turbine, structure, operations, and environment. While further equipment is needed to extract key data for predictive maintenance, value is already seen in utilizing existing data and integrating insights from different source systems. Kongsberg Digital is interested in optimizing maintenance planning and sees additional value in using DT for induction and training, as witnessed in oil and gas DT. The biggest challenge is anticipated to be the interaction between original equipment manufacturer (OEM) and operator, as well as accessing real-time high-resolution data and 3rd party sensoring. Furthermore, the interaction of the DT with the turbine and the software will be challenging. Kongsberg Digital's recommendation for academic research is to focus on data standardization and autonomy in DT.    
    \subsubsection{Kongsberg Maritime} 
    Kongsberg Maritime is part of the Kongsberg Group with a focus on marine technology. Kongsberg Maritime focuses on condition monitoring and condition-based maintenance of rotating parts and electric health, but highlights predictive capabilities as the most important technology. The goal is to prevent unexpected and long-lasting downtime for customers.
    With the sister company Kongsberg Digital, Kognitwin is Kongsberg Maritime's preferred choice as a DT platform. While there are knowledge and resources available from the sister company, more knowledge is required to build a DT. Additional limitations exist in the budget. Digitalizing the assets will require additional investment in hardware instrumentation. Kongsberg Maritime anticipates the biggest challenge to be acceptance in the industry, market, class societies, and insurance companies.
    \subsubsection{Mainstream Renewable Power} 
    Mainstream Renewable Power, which is part of the Aker group of companies, is a leading pure-play renewable energy company, with offshore and onshore wind and solar assets across global markets, including in Europe, the Americas, Africa, and Asia-Pacific. Mainstream’s goal is to establish a DT of a floating offshore wind installation to enable continuous improvements in production and O\&M optimization and lower the LCoE of operating floating wind farm assets. One important aspect is to monitor the environmental impacts of floating offshore wind farms and find effective mitigation strategies. While DT from levels 0-3 already exist in various forms in the industry but are often named differently, the real value of a new DT generation would be to achieve levels 4 and 5 which is very challenging. The biggest challenge here is the accuracy and speed of the data-driven and physical-based modeling which is needed to build valuable DTs from level 3 onward. Together with Cognite, Mainstream Renewable Power is developing a DT in the research project NextWind funded by the Californian Energy Commission.
    \subsubsection{Norconsult}
     Norconsult is the largest consulting company in Norway and has many different departments involved in wind energy topics: wind resources, construction, environment, and electric grid, to mention a few. They are mainly involved in the preconstruction phase and will typically provide advice on cost-effective solutions based on best practices. To this end, they are dependent on digital models, mainly physics based. The abstraction level of the model depends on the task at hand. To obtain more cost-efficient solutions across disciplines, Norconsult envisions individual models to propagate uncertainties associated with data and model abstraction level, from input environmental data to bankable financial analysis output. In this way, unnecessary and costly over-capacity will be removed, and risk will be managed in a coherent way. Norconsult's models tend to be computationally heavy and popular methods of propagation of uncertainties through integrated complex systems rely on “the law of big numbers” (Monte Carlo simulations). This requires a suitable abstraction level and fast codes. High license-prizing prevents collaborations, and free software supported by big governmental institutions become increasingly popular. Norconsult's suggestion for academic research is to further develop model components at different abstraction levels, making sure that these can “talk” to other components in generic, module-based integrated models. The models should also be formulated so that uncertainties can propagate through the systems where correct uncertainties are assigned to individual process. It is thus paramount that the model systems digest data in some way and preferably correct and update themselves. Clever data gathering will utilize model information to determine data parameter, location, and temporal resolution of sensors.
    \subsubsection{SNSK} 
    Store Norske Spitsbergen Kulkompani is interested in DT for hybrid energy systems in remote Arctic and Antarctic locations, where the DT is an important tool to be used in an operation center that supports local personnel. SNSK's focus is on improving component lifetime and planned maintenance through condition monitoring and \textit{what-if ?} scenario analysis. Therefore, SNSK starts to generate value through a DT from level 2 (diagnostic) onward. SNSK monitors assets through Datavaktmesteren with a real-time data stream, and autonomous decisions are made, but no predictions or \textit{what-if ?} analyses are included. 
    SNSK would like research to focus on integrating wind turbines with other power producing units and energy storage and distribution systems. 
    \subsubsection{Statkraft} 
    Statkraft is Europe’s largest renewable energy producer and a global company in energy market operations. Statkraft is interested in using DT for optimizing the O\&M of renewable energies through condition monitoring, \textit{what-if ?} scenarios, and autonomous decision-making. \textit{what-if ?} analysis is highlighted as the most interesting aspect of DT to analyze the consequences of parameter changes and asset modifications. Statkraft already has a large real-time data stream and the infrastructure to feed DT models. Statkraft is interested in all capability levels of DT but the descriptive and diagnostic are the most interesting ones. The biggest challenge is expected to be developing and distributing the knowledge of building a physical model and developing and maintaining a secure software system to run DTs. Statkraft's recommendation for academic research is to focus more on the earlier steps of realizing DT.
    \subsubsection{TotalEnergies} 
    TotalEnergies is a multi-energy company that produces and markets fuels, natural gas, and electricity. TotalEnergies wants to use DT for operation and maintenance optimization, specifically for condition monitoring. Furthermore, value is expected from utilizing the extracted data for future design improvements. It is desired to model the whole wind farm with wind, drivetrain, electricity production, structural fatigue, and mooring tension. Modeling only parts of an asset is not sufficient. The minimum DT level to extract value is descriptive (level 1), but TotalEnergies aims for prescriptive DT (level 4) to optimize maintenance and prevent downtimes by supporting decision-making for O\&M teams. TotalEnergies believes that the sensors in existing turbines are sufficient for building a DT, but sensors at other places might be required for e.g. structural integrity monitoring. TotalEnergies anticipates the biggest challenge to be the integration of all data from the wind farm into a single unified system but also highlights the value of standardization of data streams and storage. However, academic research should focus on how to combine numerical modeling and measured data, on how to limit the number of sensors and optimize their position, on how to generate value from the collected data, and on how to ease the decision process for inspection and maintenance.
    \subsubsection{Vard} 
    Vard builds ships as part of Fincantieri. Through DT technology Vard aims at improving the design and increasing the operability of vessels. Vard's focus is on the digitalization of important parts and their components, modeling the whole vessel is not required. Vard sees forecasts and \textit{what-if ?} scenario analysis during the operation and design phase as the most important features of DT. Only a few new sensors will have to be installed on the vessel, however, a platform for data streaming and remote control is under development. Vard expects the current lack of a common ontology for fully integrated windfarms as the biggest challenge. Furthermore, research should focus on developing accurate forecasts.
\subsubsection{Equinor}  
    Equinor is an international energy company committed to long-term value creation in a low-carbon future. Equinor’s portfolio of projects encompasses oil and gas, renewables, and low-carbon solutions, with ambitions to take a leading role in the energy transition and become a net-zero energy company by 2050. Equinor recognizes its strong ability to apply new technologies and digital solutions as a competitive advantage. Digital technology is a key enabler for Equinor to deliver on its ambitions. Equinor views the digital twin as a digital representation of a physical asset that fulfills the requirements of specific use cases. During the asset design and engineering phases, Equinor envisions utilizing the digital twin to effectively monitor and manage the asset development process. This involves conducting consistency checks, verifying the engineering design, and managing the master engineering data. In the asset operational phase, Equinor envisions leveraging the DT to optimize maintenance, modification, and operational performance, thereby increasing production and reducing costs associated with the assets. This entails activities such as condition monitoring, predictive maintenance, planning, continuous optimization, as well as simulation and scenario testing to explore potential scenarios. From Equinor's perspective, the concept of an overall asset DT involves a network of interconnected DTs that support different use cases, perform different functions, and involve various services (both internal and external). For Equinor, a fundamental aspect of the DT concept is the capability to evolve to follow the asset needs during its entire lifecycle. The digital representation of the asset should support the stages of asset design and engineering, and continue to evolve to meet the needs of the operation and maintenance phases, thus maximizing the asset's value potential over its entire lifespan. This is a collaborative effort that entails the involvement of multiple industry stakeholders such as OEM suppliers, EPCI contractors, IACS vendors, and service providers, who contribute to the development and evolution of the DT throughout the asset's lifecycle. For Equinor, in order to achieve the full value potential of the DT concept, it is fundamental that DT development and implementation adhere to open architectures and industry standards. This ensures data and information interoperability, facilitates standardized industry practices for data integration and machine-readable formats and enables seamless machine-to-machine communication. In addition, Equinor believes it crucial that DT solutions should ensure the preservation of completeness, accuracy, trustworthiness, and structure of the asset data and information.

    \begin{figure}
	\centering
	\includegraphics[width=\linewidth]{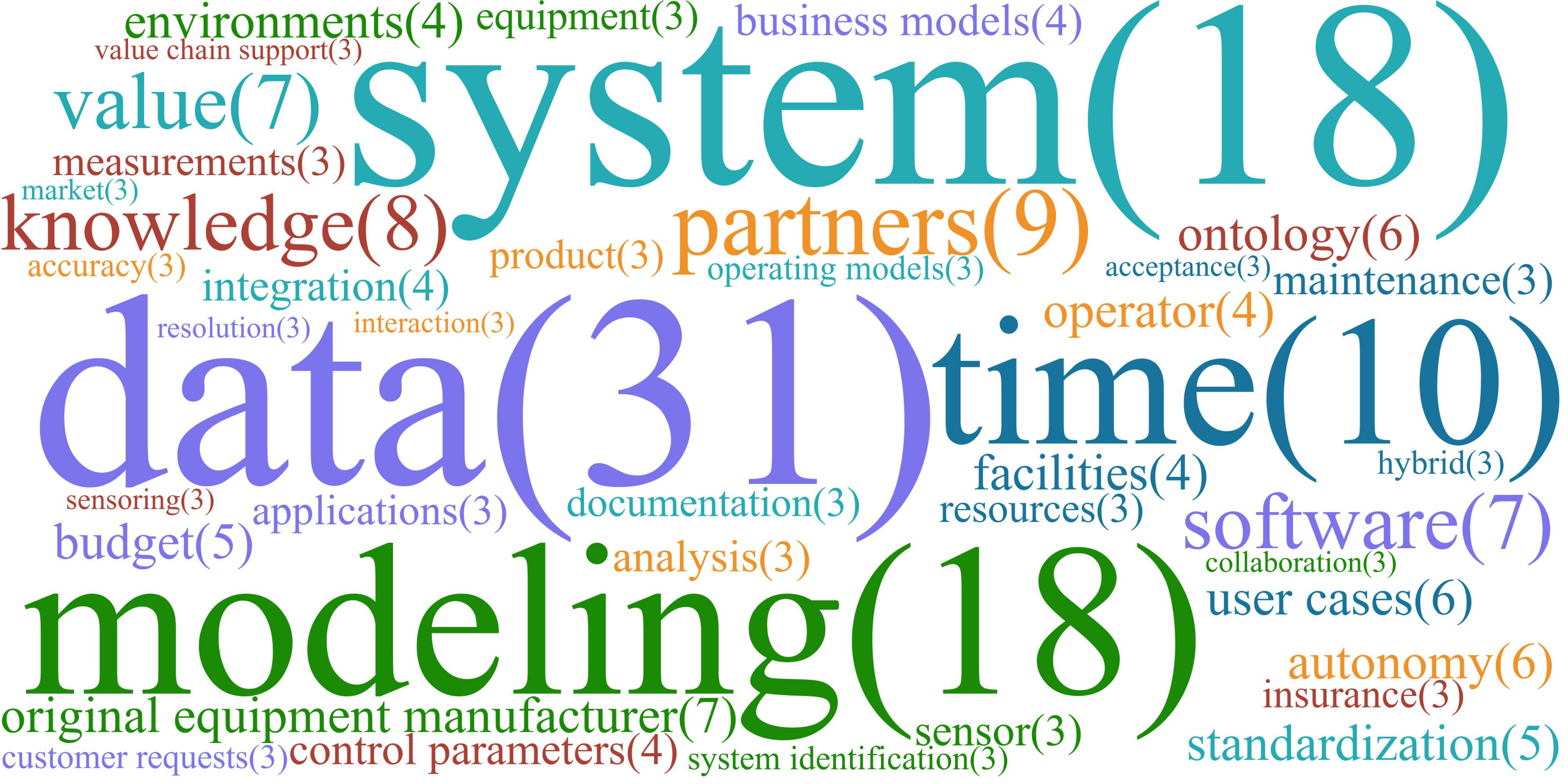}
	\caption{Word cloud with challenges of establishing digital twins that the industry partners experience, have experienced, and anticipate experiencing in the future. The cloud is compiled from the feedback throughout all three stages of the industry survey. The font size scales with a number of industry partners mentioning the same keywords.
    Data connect the digital twin with the real asset, but data quality, sparsity, security, and standardization all impact the realism of the digital twin. Especially, proprietary rights present a challenge that requires partnerships and collaboration between industry and academia as well as between operators and original equipment manufacturers.
    Spartio-temporal gaps need to be bridged by models, but complex physics-based models cannot be executed in real-time. Data-driven models, on the other hand, lack in terms of generalizability and reliability. Hybrid models can bridge the gap, but more research and validation are needed.
    Both data and models need to be integrated into a single system. 
    User cases need to demonstrate value to the industry for them to invest time and money into digital twins, and documentation and standardization are required for the industry to accept digital twins. Finally, valid business and operating models must be in place for the industry to commercialize digital twins.} 
	\label{fig:industriesperspective_challenges}
    \end{figure}
    
    \subsection{Common identified challenges}
    \label{ssec:common_challenges}
    It was realized that most of the industry partners can already start generating value from standalone and descriptive DTs. When it came to the desired features in DTs, condition monitoring was the obvious choice followed by predictive maintenance, and \textit{what-if ?} scenarios for the optimal operation of the asset. It is noteworthy that for many features the industry partners were also in favor of full autonomy if safety and loss of revenue are not of concern in the application. Only 6/14 industry partners highlighted autonomous decision-making as one of their main priorities, but 11/14 industry partners expressed general interest in autonomous decision-making. All industry partners agreed that at some point, additional sensors have to be installed for optimal efficiency of the DT, but 5/14 industry partners argued that the current installations are sufficient to build a DT upon. Furthermore, more than half of the industry partners reported that it is sufficient to model only parts of the wind farm/turbine to generate value with a DT. From the survey, the challenges (Figure\ref{fig:industriesperspective_challenges}) identified can be put into three broad categories explained in more detail below. 
    
    \subsubsection{Data}
    Developing a DT for wind energy requires real-time data exchange between the physical asset and the DT. However, various data-related challenges hinder the development of an ideal DT.
    
    \begin{itemize}
    \item \textit{Data quality:} To ensure the precise and accurate behavior of the DT, data quality is a prerequisite. However, duplicate, missing, ambiguous, inconsistent, asynchronous, and inaccurate data create obstacles in realizing an ideal DT.

    \item \textit{Data sharing:} Data collection and distribution involve multiple vendors who generate value from data and hence see data sharing as an advantage given to their competitors. It is generally a tedious job to reach a data-sharing agreement due to the involvement of multiple players from the development through operation to the decommissioning phase of wind farms. In the survey, it was realized that wind farm owners might not even have access to the data from their wind farms. Furthermore, if the data owner wants to share the data, they must revisit the original agreements with multiple players involved to ensure that no clause in the agreement is breached. This delays the timely extraction of value from the data.

    \item \textit{Big Data issues:} Data can be characterized by 10 Vs - large volume, velocity, variety, veracity, value, validity, variability, venue, vocabulary, and vagueness. This implies that data size, generation rate, type, quality, usefulness, governance, dynamic and evolving behavior, heterogeneity, and semantics can pose challenges.
    
    \item \textit{Data silo and interoperability:} DTs in the context of wind energy can be developed at the wind farm or individual component level, resulting in data silo issues. Different vendors may use different proprietary standards, data formats, and tools for acquiring and accessing data, making it difficult for other vendors to extract value from it. The data silo issues result in an incomplete overview of the asset performance, hinder collaboration, and lead to inefficiencies.
    
    \item \textit{Lack of systematic data collection:} Although many sensors are used to instrument different components of a wind farm, very little thought is put into the optimal placement of sensors. This results in either redundant data being recorded or no data being recorded.
    
    \item \textit{Sparse data:} Due to expensive instrumentation, data recorded might be very sparse in time and space, and a paradigm shift in the analysis is required to generate value from the sparse data.
    
    \item \textit{Lack of centralized expertise, compute power, and bandwidth to extract value from data:} Even when all the above data-related issues are resolved, the sheer variety in data will require human resources trained in a wide variety of expertise to generate insight from data. Furthermore, huge computing power will be required in one place.
    
    \item \textit{Data security:} Ensuring data security is a major challenge in data management for DTs.
    
    \item \textit{Choice for data management solutions:} Choosing the right data management solution is crucial for the successful implementation of DTs for wind energy.
    
    \item \textit{Talent gap:} The current lack of personnel trained in developing DTs poses a significant challenge for the wind energy industry to adopt DT technology effectively.
    \end{itemize}
   
\subsubsection{Modeling}

Data collected from an asset has a sparse resolution in space and time. Moreover, they are available only for past and current situations, thus posing challenges in instilling physical realism in a DT. Models can help improve the resolution of the data and, at the same time, provide insight into the future state of the asset. However, before the full potential of any modeling approach can be exploited in the context of DTs, some model-related challenges need to be resolved. Based on the survey, we present some of the most desirable characteristics of any modeling approach.

\begin{itemize}
\item \textit{Accuracy and certainty:} Accuracy refers to a model's ability to model phenomena of interest as closely as possible so that the observed state of the asset is indistinguishable from the modeled state. However, accuracy alone is not sufficient. For the models to be used confidently, uncertainty quantification is also important. The model's accuracy and certainty can suffer in the absence of a complete understanding of the underlying cause driving a phenomenon, simplifications arising from the need to make the models computationally efficient, uncertainty in the physical parameters/input to the models, and numerical errors. While in standalone, descriptive, and diagnostic DTs, reliable measured data can help improve the model's accuracy and certainty, for predictive, prescriptive, and autonomous DTs, a lack of observed data complicates things further.

\item \textit{Computational efficiency:} This is the property of a model that relates to the number of computational resources (compute time and infrastructure) required to produce accurate results with quantified uncertainty. One can afford to run computationally demanding models in a standalone DT. However, computational efficiency is paramount at other capability levels to keep the DT in sync with the asset. Also, the need for real-time modeling in predictive DTs and the need for better than real-time modeling for control and optimization in prescriptive and autonomous DTs makes computational efficiency even more desirable. Computationally efficient models will also be desirable when running on standalone hardware like a portable virtual reality set with limited computing power.

\item \textit{Generalizability:} A model's generalizability refers to its ability to solve a wide variety of problems without any problem-specific fine-tuning. An asset encounters a wide array of scenarios during its lifetime. Some of these events might be rare, while others are persistent. Since the DT is required to model all these phenomena to be in sync with the asset, the models used should be able to generalize its learning without any compromise on its accuracy and computational efficiency.

\item \textit{Ever-evolving:} It refers to a model's ability to learn and evolve every time a previously unseen scenario is encountered. A model that is specialized for a narrow range of scenarios likely loses accuracy when encountering new situations. This is especially relevant in cases of exceptional circumstances like impending hardware faults.

\item \textit{Interpretability and trustworthiness:} A model is considered interpretable when humans can readily understand the reasoning behind predictions and decisions made by the model. Trust and transparency of DT behavior have been identified as one of the biggest challenges in building a DT. Monitoring and controlling high-value assets like wind farms with DTs requires trust in the DT and therefore in the chosen models. With an interpretable model, analysis output can be backtraced all the way to the relevant sensor input. A black-box model might detect a fault in a turbine, but it requires an interpretable model to pinpoint the exact cause and plan countermeasures.
    
    \item \textit{Robustness and stability:} The robustness and stability of a model are critical factors in determining its reliability. A model that is robust and stable produces consistent results even when subjected to perturbations. In contrast, a model that is not robust may show significant deterioration when faced with noisy inputs, while an unstable model may fail for certain modes. Therefore, the robustness and stability of a model are essential requirements for ensuring its reliability.
    \end{itemize}
         
    Existing approaches can typically be grouped into three categories: physics-based, data-driven, and hybrid modeling and analysis. Each approach has its own benefits and weaknesses. High-fidelity physics-based models (PBMs) are typically associated with high computational effort, while data-driven models often lack interpretability. In the survey, it became evident that the industry needs to generate value from data while still being able to utilize existing knowledge to the maximum extent possible. To this end, multiple companies argued in favor of hybrid analysis and modeling to compensate for the weaknesses of both physics-based and data-driven modeling and recommended academic research to focus on this area.
    
    In Section~\ref{ssec:modeling} we give examples of physics-based, data-driven, and hybrid analysis and modeling techniques, their application within wind energy, and their strength and weaknesses with respect to each other and with respect to their applicability to DT.

        \subsubsection{Industrial acceptance and trust for digital twins:}
        In order for DTs to be successful, it is important to gain acceptance and trust from both the industry and the general public. This can be achieved by raising awareness about DT concepts and demonstrating the value that this technology can generate. User cases are a valuable tool in supporting trust-building.
        
        However, there are several challenges that must be addressed in order to gain industrial acceptance and trust for DTs:
        
        \begin{itemize}
        \item \textit{Little consensus on the meaning of digital twin:} There is currently little consensus on what exactly constitutes a DT, and this lack of a clear definition can lead to confusion and mistrust among industry professionals and the general public. It is important to establish a common understanding of the term and its applications in order to build trust and acceptance.
        \item \textit{Little awareness about the values that digital twins can generate:} Many industry professionals and members of the public are not aware of the potential benefits that DTs can offer, such as improved efficiency, reduced downtime, and increased safety. It is important to educate stakeholders on the value of DTs in order to build trust and acceptance.
        \item \textit{Technology readiness level is insufficient. Sustainability readiness level:} Despite the potential benefits of DTs, the technology readiness level (TRL) is not yet sufficient for widespread adoption. In addition, there is a lack of focus on sustainability readiness level (SRL) which is necessary to ensure that the technology is environmentally and socially responsible. Addressing these concerns is crucial in gaining industry acceptance and trust.
        \item \textit{Lack of appropriate business models:} There is currently a lack of appropriate business models for DTs, which can make it difficult for companies to justify the investment. The development of viable business models will be key to gaining industry acceptance and trust.
        \item \textit{Security and privacy:} The use of DTs raises concerns about security and privacy, especially in critical infrastructure such as energy, transportation, and healthcare. It is important to develop robust security and privacy protocols in order to gain industry acceptance and trust.
        \item \textit{Enabling twin projection:} One of the key benefits of DTs is the ability to project and simulate the behavior of physical assets. However, this requires accurate and reliable data, which may not always be available. In order to gain industry acceptance and trust, it is important to develop reliable methods for enabling twin projection.
        \end{itemize}

        In summary, DTs have the potential to revolutionize many industries, but in order to gain acceptance and trust, it is important to address the challenges related to consensus, awareness, technology readiness, business models, security and privacy, and enabling twin projection.

    
    \section{Existing state of the art to address the challenges and industry needs}
    \label{sec:sota}
    In the previous section, several challenges were discussed. In this section, we address each challenge with potential solutions.
    \subsection{Standards and asset information model} 
    Standards are essential for sharing, scaling, and reusing software structures as complex as DTs.  It can be expected that multiple companies are involved in a DT over its lifecycle: component manufacturers can use data from previous DT to improve the component design and may provide a digital representation of that component. OEMs use their overall Reference Designation Structure (RDS) together with operational data from previous designs to provide an updated and improved digital twin as basis for new design, sizing construction, and logistics. Maintenance crews might be employed by different companies. The energy distribution industry benefits from increased reliability and may use forecasts to stabilize the grid. Other companies might be involved during decommissioning. In addition, all involved companies can generate value from the data recorded during the operation. Finally, industries might be interested in collaborations to exchange data and DT structures for mutual benefits. 
    
    In this regard, the importance of a standardized Asset Information Model (AIM) cannot be more stressed upon. An AIM is a digital representation of an asset that contains all relevant information throughout its lifecycle, from design and construction to operation and maintenance. It is a structured collection of data that captures all the necessary information about an asset, including its physical and functional characteristics, operational requirements, and maintenance history. It also contains a wide range of information about an asset, including its physical components, technical specifications, maintenance schedules, and operating procedures. It may also include information about the environment in which the asset operates, such as weather patterns and local regulations. The purpose of an AIM is to provide a comprehensive and integrated view of an asset, enabling better decision-making and more efficient operation and maintenance. By capturing all relevant data in a single location, an AIM can help asset owners and operators to optimize asset performance, reduce maintenance costs, and extend asset lifespan. However, despite the obvious value of the standardization of AIM, very little work has been done so far in the context of wind energy. We in the following section present some preliminary work done in this direction by DNV. The structure and standards proposed in the following paragraphs have been strongly influenced by DNV's FlowSite proposal in the NorthWind project.

    \paragraph{Reference Designation System} 
    RDS provides a system for naming and structuring asset and workflow components. ISO81346-10:2018, also called RDS-PP, is an RDS for power systems~\cite{RDS_PS81346_10ti8}. RDSs are already used in wind farms \cite{VGB2022rpa}.
    
    \paragraph{Reference Data Library}
    A Reference Data Library (RDL) contains, for the most part, static or slow-changing data relevant to the system. As an example, it could include conversion factors between units, physical constants, or country codes. The Reference Data Library ISO 15926-4 was primarily intended for oil and gas but is also used for other areas. 
    
    \paragraph{Semantic data model}
    In a semantic data model, the meaning of instances is described. There is a working draft for part 14 of ISO 15926. It presents an "Industrial top-level ontology" based on five years of experience from a Norwegian group of oil and gas companies~\cite{Walther20ias}.
    
    \paragraph{Information exchange and application interfaces}
    Some standards relevant here include ISO 10303 for Product data and exchange, which is also known as STEP, Standard for the exchange of product model data. Included here is, e.g. CAD model exchange~\cite{Pratt2001iti}. 
    In IEC 61970-3 the Common Information Model (CIM) for the "semantics of information exchange" is presented \cite{Santodomingo2016i6f}. It is usable for application program interfaces for energy management systems.
    IEC 61968 can be used to define information exchange between electrical distribution systems but appears to be still in development. 
    
    These standards can be partially adapted, but it can be expected that extensions will have to be done for DTs, especially with respect to the application interface.
    
    \paragraph{Digital Twin environment}
    Simulations for wind farms have been performed before, but with DT technology, they need to handle large amounts of input data and perform simulations in real-time. It has to be assessed if existing platforms can be adapted to this new situation or if specific DT environments have to be created. An example of such a platform is the Open Simulation Platform, which is being built by DNV, Kongsberg, SINTEF, and NTNU~\cite{osp}.
    During the survey, Microsoft Azure DTs have been mentioned multiple times for hosting DTs. However, there are more specific products, such as Kongsberg's Kognitwin~\cite{KongsbergDigitalk} or 4Subsea's 4insight \cite{4subsea4}. %
    
    \paragraph{Required Standards}
    While the above-mentioned existing standards provide a basis, DTs will require additional standards. This includes standards for data compression, standardized connections between data, models, analysis tools, and interface, and standards for interaction between DTs. Basing these standards on existing ones will improve industrial acceptance.
    
    \subsection{Data acquisition, communication, sharing, archival}\label{ssec:data_generation}
    In DTs, data from many sources are combined to build an as-realistic-as-possible virtual representation of the physical asset and its environment. Some information like architecture and design is usually static and can be inserted manually. Most information such as wind speed, temperature, wave height, and loads are changing constantly and have to be updated in real-time through sensor measurements. The collected data are then compressed, augmented, and/or analyzed with physics-based and/or data-driven models. PBMs make use of domain knowledge, such as the fundamental laws of nature. Data-driven models, on the other hand, are built and trained with a large amount of data collected from similar assets in the past. In this subsection, we address both real-time and historical data. We discuss modeling techniques in Section~\ref{ssec:modeling}. 
        
        \subsubsection{Sensors} 
        While the industry in general agreed that at some point additional sensors will have to be installed, some companies argued that the currently installed sensors can already be used to build DTs, thereby reducing the monetary entry threshold of DTs for wind farms.
        The variety of sensors usable in wind farms is enormous. Heat can be measured through temperature sensors and indirectly provides information on friction on gears and bearings. Acceleration sensors allow vibration measurements that can provide information on component wear. Microphones measure the vibration through air. Tower loads and strains are measurable with corresponding sensors. Oil quality can be monitored, air and water corrosion can be tracked through air and water concentration measurements, and precipitation measurements contribute to blade damage estimation. Specifically interesting for wind fields and therefore farm control are LIDAR systems (light detection and ranging~\cite{Smith2014nit}) sensors. LIDARs use lasers and the Doppler effect to measure wind speed. They can be mounted on the nacelle~\cite{Trabucchi2017nbl} or on separate buoys~\cite{Gottschall2014rac} and can be used to measure the speed of the incoming wind~\cite{Smith2014nit} as well as to measure wake effects~\cite{Shin2022eso}. This helps to model an accurate wind field around and within the wind farm, which can then be used to improve the accuracy of the wind impact on each turbine. With the success of machine learning in image recognition, cameras provide a valuable measurement technique. In~\cite{Sundby2021gcd} a combination of dynamic mode decomposition and two NN-based models are used to identify the movement and rotation of objects in recorded images. The technique can be generalized to spot environmental changes in the wind farm and its vicinity, be it boats or buoys in offshore farms or agriculture or forestry in onshore parks. 
    
        \subsubsection{Optimal sensor placement}
        The sensor positions are the basis for accurate data acquisition. Through a sophisticated Design of Experiment (DoE), the number of sensors can be minimized, while the measured information is maximized. In a holistic level 1+ DT of a wind farm, measurements for, e.g., fatigue loads at towers, foundations, blades, and drivetrains have to be performed. In~\cite{Mehrjoo2022osp} the Sequential Sensor Placement algorithm is used by minimizing the information entropy of the relevant quantities. They not only maximize the accuracy but also consider the installation cost of sensor positions to determine the strain time history and tower-, jacket support structure- and soil-stiffness of an offshore wind turbine. The algorithm has to be used differently for parameter estimation and stain estimation. In \cite{Pichika2022osp} the authors investigate the optimal positioning of sensors for fault analyses in turbine gearboxes based on statistical features and data-driven methods. Their algorithm is tested on a lab scale and they are able to reduce the number and increase the fault detection accuracy of sensors, but they note that their method needs further investigation on different data sets. In \cite{Schulze2016osp} the optimal placement of a given number of acceleration sensors for the static modal analysis of the blades is determined using a multibody approach. Three different algorithms were tested on a 3MW onshore prototype by using the modal shape matrix and testing the results with the Auto Modal Assurance Criterion. A genetic algorithm with a weighted off-diagonal criterion identified the sensor positions with the lowest linear dependencies between measured modes.
        
        \subsubsection{Reducing data size}
        Through the number of sensors and high measurement frequencies, the data stream between physical assets and DT can be rather large. Compression and decompression of the data can help to reduce the required bandwidth and storage space. Compression techniques are applied before the data transmission is performed and decompression is used before using the data. Other techniques reduce the amount of sensor data transmitted or the measurement frequency. Commonly known lossless compression algorithms making use of the file string include LZMA (e.g. .7z), Deflate (e.g. .zip), or BWT. They use features like, e.g. identifying identical strings and replacing them with pointers, or rearranging the coding (length) for symbols according to their frequency. LZMA (Lempel–Ziv–Markov chain algorithm), Deflate, BZIP2, and GRIB2 (General Regularly distributed Information in Binary form: Edition 2) are tested for wind power and wind speed data in~\cite{Louie2012lco}, where they are combined with different preprocessing schemes. Principal Component Analysis (PCA) enables the reduction of the dimensionality of data to the components with the highest variance. 
        
        Autoencoders perform NN-based compression. The input and output layers have the size of the data, but at least one of the intermediate layers is smaller in size. The NN is trained to output the input data. Afterward, the NN is split at the smallest layer, so that the first part can be used as an encoder and the latter one as a decoder. Generative Adversarial Networks are another type of NNs. A Generator tries to reconstruct the original data, while a Discriminator tries to separate the reconstructed data from the original data. By training both against each other, even data with unknown compression algorithms can be restored. The drawback of NN-based methods is that they only perform well on the parameter space on which they are trained. If sensors record unexpected data for example from damage to the farm, they will not be able to recover the data after compression. However, this reconstruction error can be monitored by performing decompression on the farm and comparing the accuracy of the data. Fault detection is done with Autoencoders in~\cite{Roelofs2021aba} and with Conditioned Variational Autoencoders in~\cite{Mylonas2021cva}.
        
        Compressed Sensing can be used for sparse signals, i.e. signals for which there is a domain so that the signal is only represented by a few non-zero components, to recover the original signal from a reduced frequency of data points, allowing for reduction of the sampling frequency and thereby the data size. This can be combined with e.g. a fast Fourier transformation for periodic signals. In~\cite{Du2020ees} compressed sensing is utilized to reduce the amount of transmitted data by predicting sensor data and only requesting sensor updates when a category of sensor data differs from the predicted category. Under the assumption of frequency compressibility of the periodic impulsive component,~\cite{Du2017csb} applies compressed sensing for the wind turbine gearbox to recover the impulsive features from fewer data points than conventionally required by the Shannon sampling theorem. 
        
        Virtual sensing is a technique used to infer the value that a sensor would measure from data recorded with other physical sensors. Physical sensors are therefore not required at points of interest but can be placed at positions where they gain maximum information. This not only allows for reducing the total amount of sensors and inferring values at places where sensors cannot be installed but also for inferring unmeasured quantities. Note that a physically realistic model is required to enable virtual sensing. The authors in ~\cite{Tarpo2022ddv} use PCA for virtual sensing of strain estimations on an offshore tower using temporary sensors for model calibration. A large number of compression approaches and use cases outside of the wind sector are listed in, e.g.,~\cite{Jayasankar2021aso}.
        
        \subsubsection{Data security}
        The gathered data has to be secure both in terms of validity and privacy. Blockchain technology has applications not only in cyber-currency. In a blockchain, data are chained together in blocks through hashes. Each data block contains the hash of the previous block, data, and a new hash generated from both. If data or previous hash in a block is manipulated, it will alter the hash of that block. Modifying a block would require modification of all the following blocks to remain undetected.
        In~\cite{Wang2021dow} the prospects of using blockchain to build a safe wind farm information system are investigated.
        
        \subsubsection{Data sharing}
        Huge amounts of data are generated in every wind farm. However, the amount of openly available data is puny~\cite{Kusiak2016rsd}.
        Especially data from the Supervisory Control and Data Acquisition (SCADA) system is not readily available~\cite{Leahy2019iwd}.
        This is partially attributed to the way proprietary rights are handled in wind farms~\cite{vanKuik2016ltr}. This became especially evident in the industry survey conducted here. It was commented on multiple times that the proprietary rights of the OEMs are causing problems even for the operators of the wind farms themselves. It was hinted that the OEMs are keeping the proprietary rights to secure their designs and expand their market to analysis and operating software. This heavily influences both the development and application of DT and related technologies such as predictive maintenance. The future will show whether DT will cause increased data sharing or if OEMs will be the only ones offering DT for their assets.
           
    \subsection{Modeling and Analysis}
    \label{ssec:modeling}
       \subsubsection{Physics-based modeling (PBM)}
        \begin{figure}[!htb]
        \begin{subfigure}{\linewidth}
        \centering
        \includegraphics[width=\linewidth]{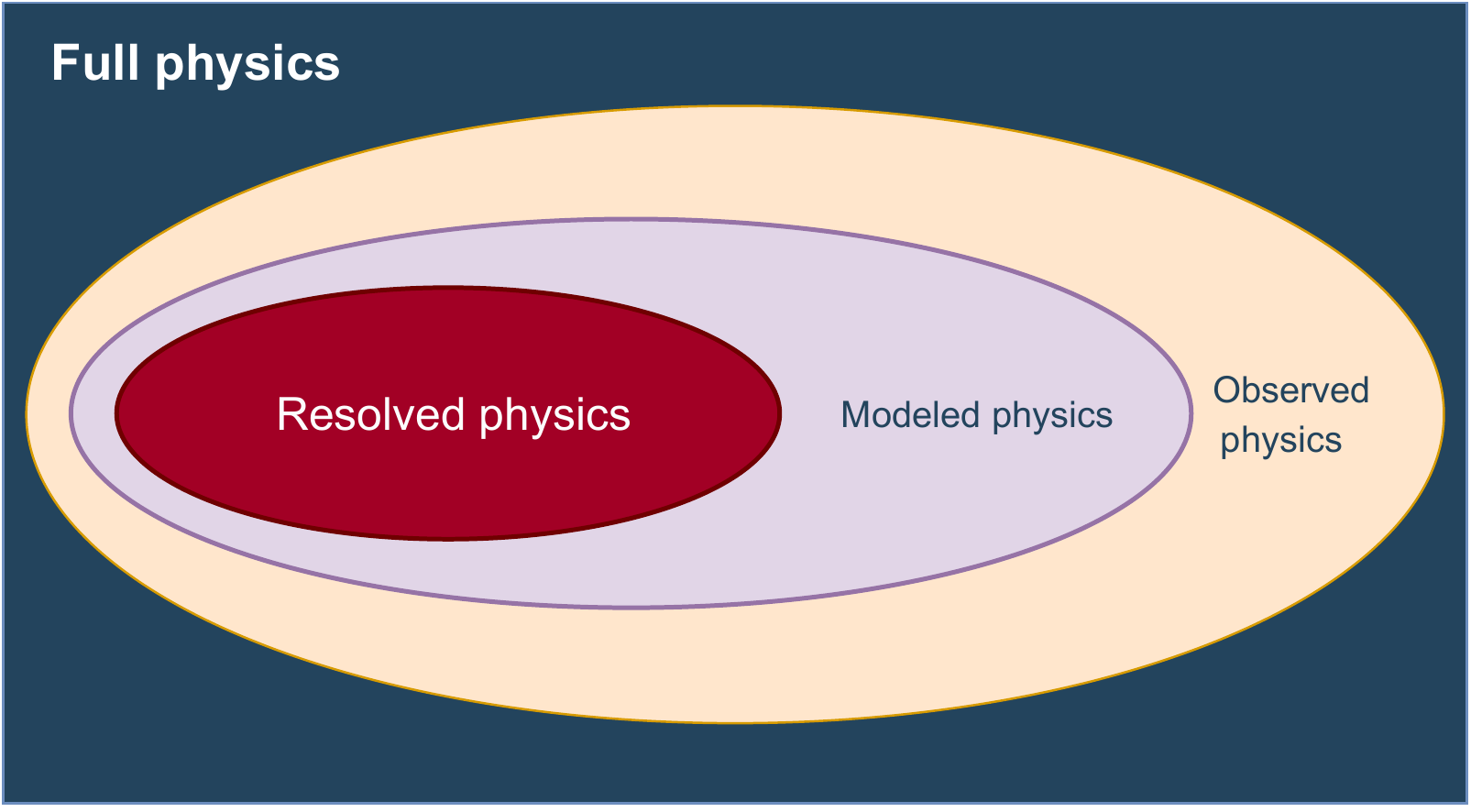}
        \caption{PBM resolves only a part of the physics} \vspace{0.5cm} 
        \label{subfig:pbmschematic}
        \end{subfigure}
        \begin{subfigure}{\linewidth}
	    \centering
	 \includegraphics[width=\linewidth]{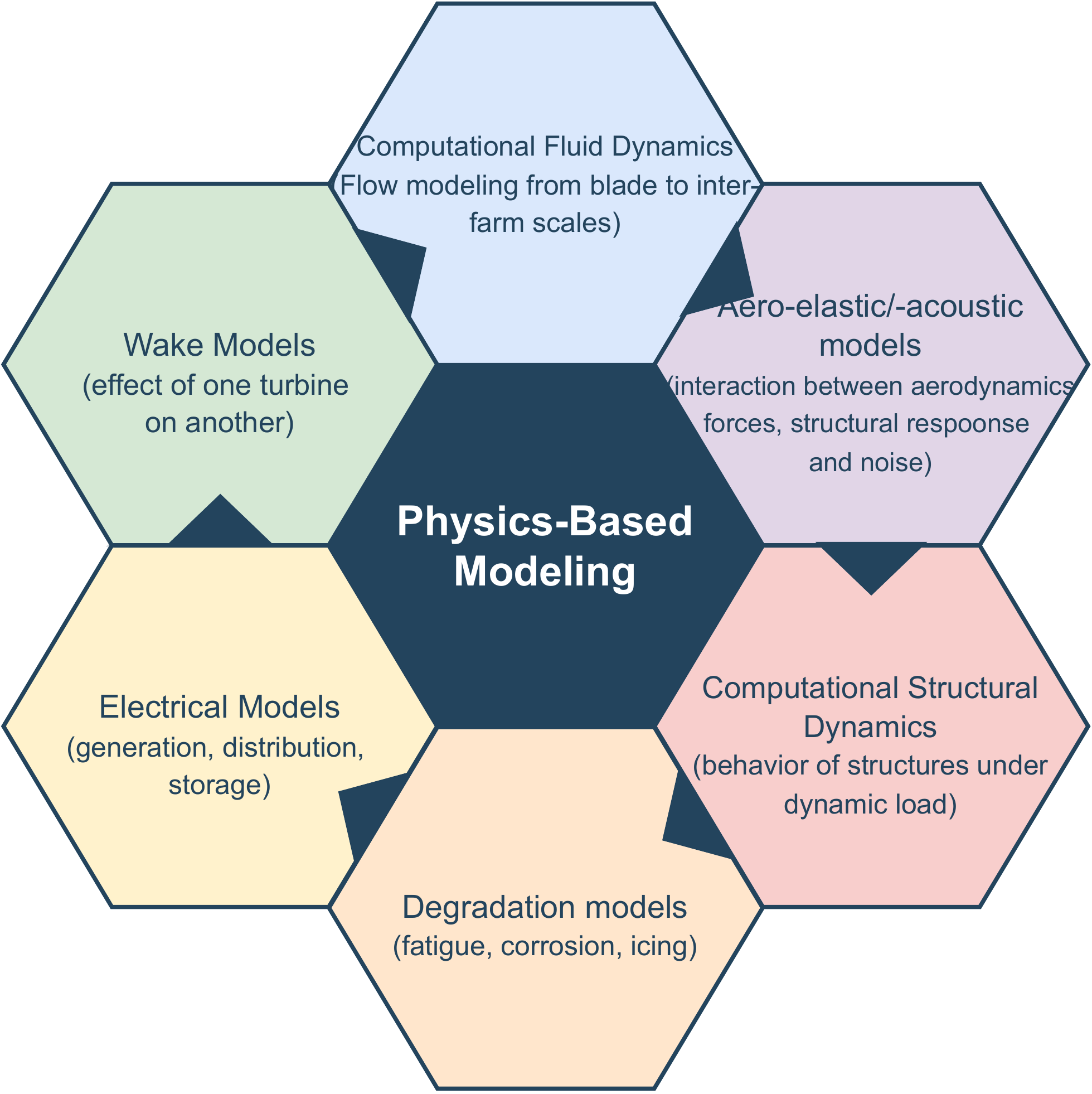}
	    \caption{Six broad categories of PBM} 
	    \label{subfig:pbmoverview}
        \end{subfigure}
        \caption{Physics-based modeling}
        \label{fig:PBM}
        \end{figure}

        This approach (Figure~\ref{subfig:pbmschematic}) involves careful observation of a physical phenomenon of interest, the development of its partial understanding, the expression of the understanding in the form of mathematical equations, and ultimately the solution of these equations. A PBM is a representation of the governing laws of nature. These laws of nature are typically defined in the form of conservation and constitutive laws, often based on theoretical development and experimental validation. They are most often represented by systems of differential equations that are approximated by numerical methods and solved on computers. Wigner~\cite{Wigner1960tue} states that PBMing is powerful and effective because it gives us a predictive window into the future based on understanding, \cite{Willcox2021tio}.
        Categories of PBMs that are relevant for wind energy, categorized (Figure~\ref{subfig:pbmoverview}) by their areas of application:
    \begin{itemize}
        \item Computational Fluid Dynamics (CFD): CFD is a branch of fluid mechanics that uses numerical analysis and algorithms to simulate and analyze the behavior of fluids. In wind energy, CFD models can be used to study the flow of air around wind turbines and predict their aerodynamic performance.      
        \item Computational Structural Dynamics (CSD): CSD models are used to study the behavior of structures under dynamic loads. In wind energy, CSD models can be used to study the response of wind turbines to turbulent wind conditions, such as gusts and eddies.
        \item Aeroelasticity models: These models are used to study the interaction between aerodynamic forces and the structural response of a system. 
        In \cite{Wiens2021hso}, a holistic simulation of wind turbines with fully aero-elastic and electrical model is presented. A detailed review of aeroelasticity in the context of a wind turbine is provided in \cite{Zhang2011roa}.
        In wind energy, aeroelasticity models can be used to study the dynamic behavior of wind turbines, including blade deflection and tower vibration. 
        \item Wake models: Wake models are used to study the behavior of the wind wake downstream of a wind turbine~\cite{Gocmen2016wtw,Pawar2022tmf, Martinez_Tossas2022nio}. In wind energy, wake models can be used to predict the impact of wind turbines on the surrounding environment, such as the effects of wind turbines on nearby wind farms and on the local wind resource.
        \item Electrical models: Electrical models \cite{Kocewiak2019mow, Gustavsen2017vtv} are used to study the electrical behavior of wind turbines, including the generation, distribution, and storage of power. In wind energy, electrical models can be used to optimize the performance of wind turbine generators and to study the integration of wind energy into the electrical grid.
        \item Degradation models: Degradation models (eg. \cite{Li2019rao, Bai2017mfd}) are used to study the long-term performance of systems, including the effects of wear and tear, aging, and environmental factors on the system's performance. In wind energy, degradation models can be used to study the aging and deterioration of wind turbine components, such as the blades, gearbox, and bearings.   
    \end{itemize}
        
        \paragraph{PBM in Standalone DT}
        Standalone DTs are disconnected from the physical wind farm. They do not need to be evaluated online, and therefore the requirements for computational efficiency are less strict than in other capability levels. 
        PBM can be used in standalone DT in the design and planning phases.
        One focus during the design stage is blade optimization. 
        In \cite{HoangQuan2021wtb} the turbine blade design is being optimized with CFD modeling software. The authors in \cite{Siddiqui2019nio} use the Reynolds averaged Navier Stokes technique to investigate geometric approximations of blade segments on the aerodynamic performance of a wind turbine. In \cite{Bazilevs2012ifs} the authors investigate fluid-structure interaction using isogeometric analysis and non-matching fluid-structure interface discretization on an offshore reference wind turbine rotor.
        Another important design aspect is the farm layout, which depends on local wind resources and wake effects.
        In \cite{Antonini2020odo} CFD methods are used to optimize the design of a wind farm in complex terrain. \cite{Richmond2019eoa} validates an offshore farm CFD model against data from an operational site. The wake growth rate is investigated in \cite{Vahidi2022apb} assuming a Gaussian wake model.

        \paragraph{PBM in Descriptive DT}
        The descriptive digital twin describes the current state of the wind farm. Even with perfect initial conditions, any PBM alone will inevitably drift away from the real state over time due to approximations and external influences. Data, on the other hand, are sparse in space and time and will not describe the wind farm and its environment in sufficient detail without models to interpolate in space and time and to infer parameters that cannot be measured directly.
        Therefore, both data and models need to be combined. Deviations between PBM and reality can be alleviated by measuring data and using them as input to the PBM for boundary conditions and re-initialization of the PBM. This strategy is already being used, for example, to nest wind flow models into meteorological forecasts \cite{Tabib2021anm}. 
        A critical condition for using PBM in descriptive and higher capability levels is the execution speed. Models that cannot be evaluated in real-time cannot be used for digital twins above the standalone level, since they cannot keep up to date with the asset. Approximations can reduce computational efficiency at the cost of accuracy. When choosing PBMs for digital twin applications, a tradeoff between speed and detail is imperative. 
        
        \paragraph{PBM in Diagnostic DT}
        Using the real-time data and spatiotemporal resolution-enhancing models from the descriptive DT, the condition of a wind farm can be monitored and diagnosed. In the context of descriptive DT, PBMs are especially interesting for degradation estimation for all wind turbine components from blades and bearings over the drivetrain, gearbox, and generator to the tower and foundation/substructure/floater. The authors of \cite{Gondle2021eow} explore the feasibility of a novel low-cost mechanical displacement indicator to continuously monitor tower movement relative to the foundation base and suggest that a combination of measurements and FEM would allow identifying foundation issues. In \cite{Montesano2016doa} a multiscale progressive damage model was implemented by combining computational micromechanics and continuum damage mechanics in a FEM to predict subcritical microscopic damage evolution and stiffness degradation in turbine blades. In \cite{Moghadam2022ocm} a DT condition monitoring approach for drivetrains is implemented by estimating the RUL through online measurements and fatigue damage estimation. It is clear that due to the inherent weaknesses associated with the ignored physics and uncertainty in input parameters, their diagnostic characteristics cannot be relied upon.

        \paragraph{PBM in Predictive DT}
        The predictive DT can include predictions of wind, weather, power output, turbine motion, temperature, loads, fatigue, and other parameters whose prediction either brings direct value to the user or can be used as input to estimate other useful quantities. \cite{Wang2011aro} reviews wind power forecasting models across multiple wind farms and countries. In \cite{Rasheed2014amw} a multiscale wind model is used to predict wind flow on wind farms. In \cite{Stadtmann2023sda}, the meteorological predictions of the model are integrated into a DT to infer wind turbine and farm power production and explain the predictions in a virtual reality interface.

        \paragraph{PBM in Prescriptive DT}
        On the prescriptive level, the demands on the execution time of integrated models become even stronger, since many \textit{what-if ?} scenarios must be explored simultaneously to obtain uncertainty estimates and provide optimal recommendations to the user. Even with parallelization, computational efficiency remains a challenge for PBMs. However, assuming sufficient speedup, models mentioned in earlier sections can be used as input for optimization and uncertainty estimates. The prescriptive digital twin can be used for decision support in scenarios where human operators have enough time to compare the value of the description with their own domain knowledge and react. Maintenance scheduling is one such example, where degradation and damage propagation models are required to analyze the risk and reward of timely or delayed maintenance while taking into account weather conditions determined to identify optimal maintenance windows. Therefore, fast diagnostic and predictive models are required as input into the scenario analysis of the prescriptive component. It is worth noting that the tools that enable standalone, descriptive, diagnostic, and predictive DT can be used in a presvcriptive setting too. 
        
        \paragraph{PBM in Autonomous DT}
        The autonomous DT can be used for applications that require continuous optimization or where human operators do not have enough time to react, and for applications that require continuous optimization. In some cases, model-free control can be sufficient, but reliability in complex scenarios argues in favor of model-based controllers. Like the prescriptive DT, fast and reliable models are needed as input for these controllers in order to optimize the control process.
       
        Despite their utility, due to the partial understanding and numerous assumptions along the steps from observation to the solution of the equations, a large portion of the important governing physics gets ignored in a PBM approach. Even the applicability of high-fidelity simulators with minimal assumptions has so far been limited to the offline design phase only. Despite this major drawback, what makes these models attractive are sound foundations from first principles, interpretability, generalizability, and the existence of robust theories for the analysis of stability and uncertainty. Unfortunately, most of the accurate PBMs are generally computationally expensive, do not automatically adapt to new scenarios, and can be susceptible to numerical instabilities.
        
        \subsubsection{Data-driven modeling}
        \begin{figure}[!htb]
        \begin{subfigure}{\linewidth}
        \centering
        \includegraphics[width=\linewidth]{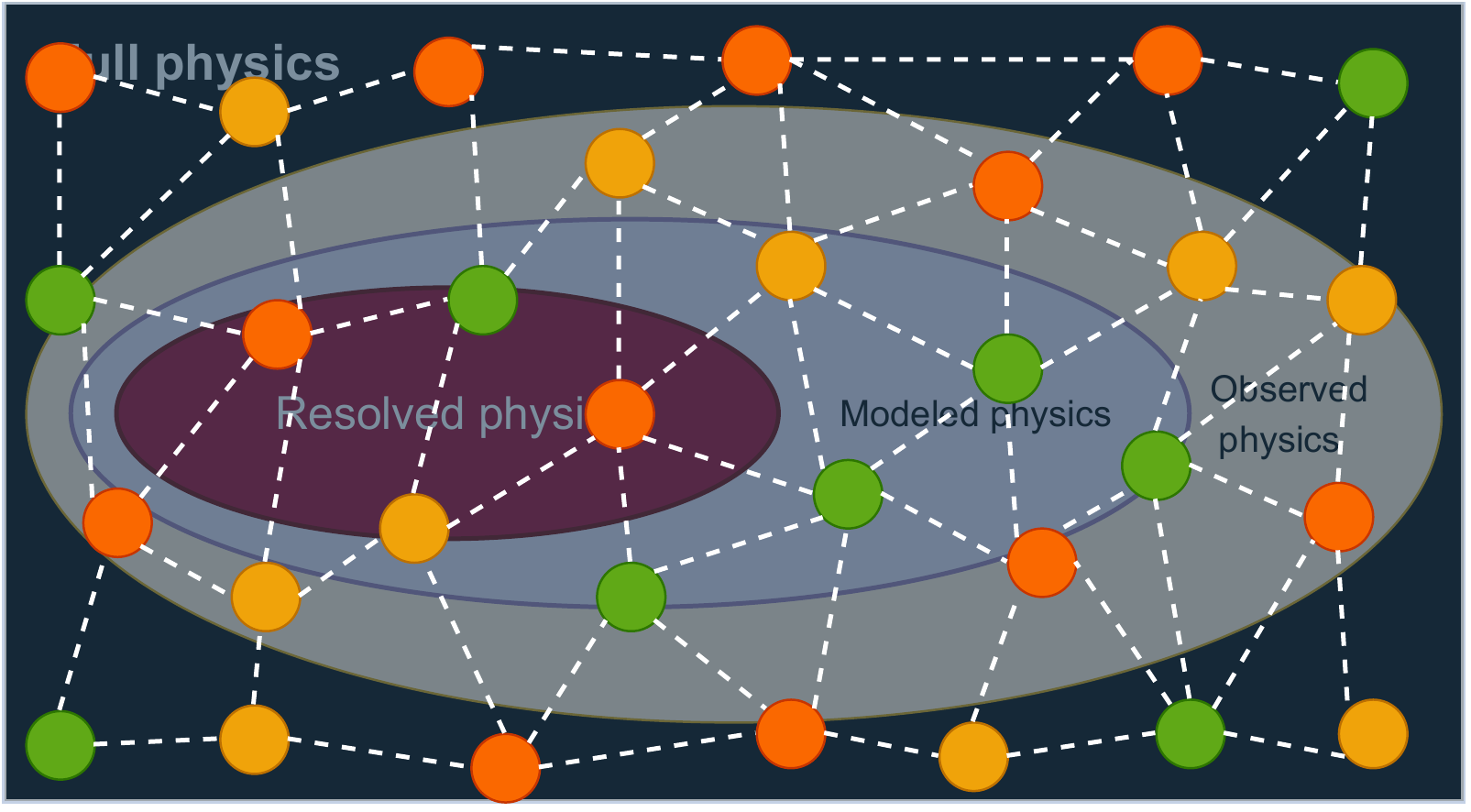}
        \caption{DDM attempts end-to-end modeling of both known/modelled and unknown/unmodelled physics}
        \label{subfig:ddmschematic} \vspace{0.5cm}
        \end{subfigure}
        \begin{subfigure}{\linewidth}
	    \centering
	    \includegraphics[width=\linewidth]{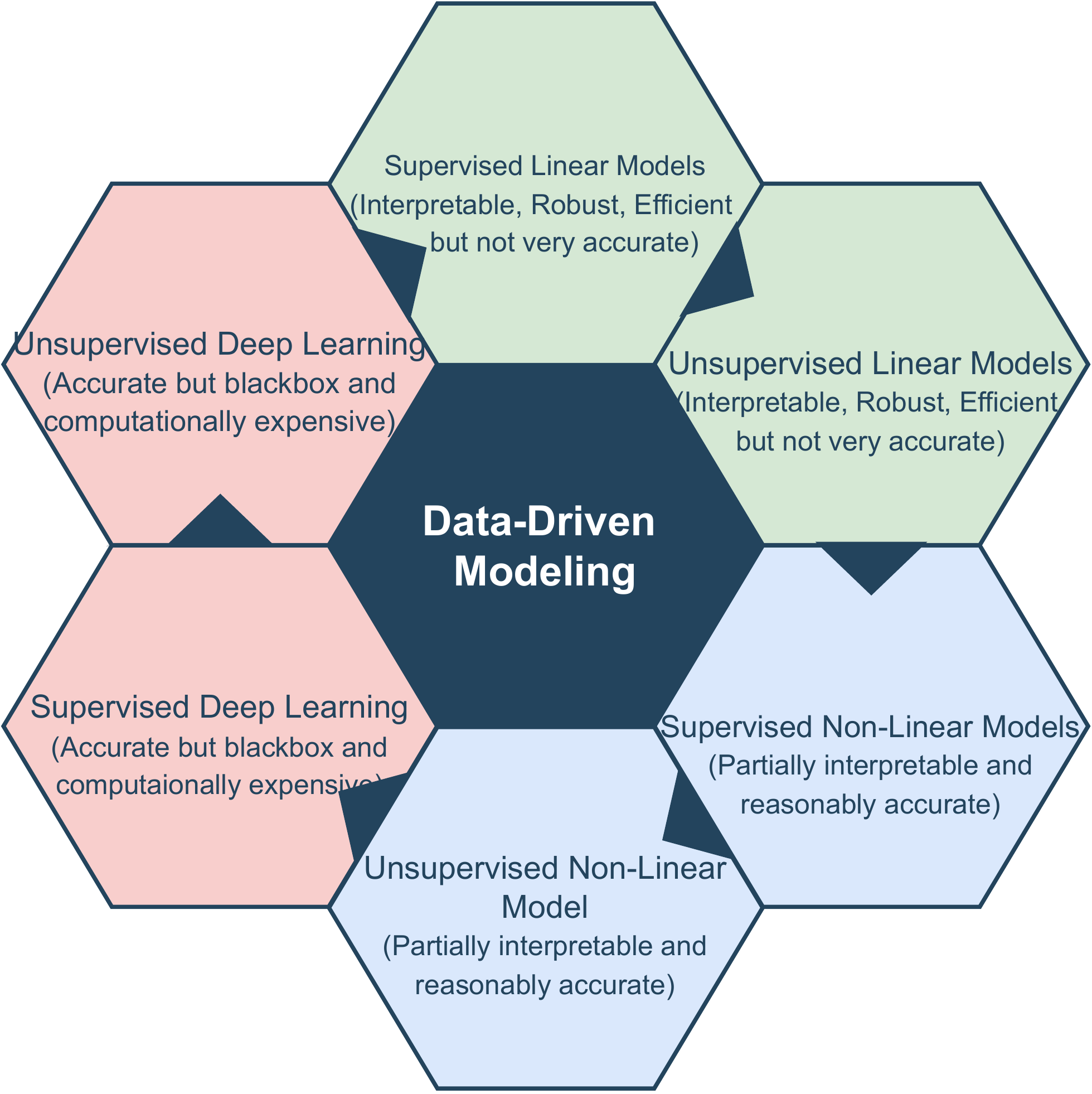}
	    \caption{Six broad categories of DDMs} 
	    \label{subfig:ddmOverview}
        \end{subfigure}
        \caption{Data-driven modeling.}
        \label{fig:DDM}
        \end{figure}

        With the abundant supply of big data, open-source cutting-edge and easy-to-use machine learning libraries, cheap computational infrastructure, and high-quality, readily available training resources, data-driven modeling (Figure~\ref{subfig:ddmschematic}) has become very popular. Compared to the PBM approach, these models thrive on the assumption that data are a manifestation of both known and unknown physics and hence when trained with an ample amount of data, the data-driven models will learn the full physics on their own. This approach, involving, in particular, deep learning, has started achieving human-level performance in several tasks that were until recently considered impossible for computers. The data-driven models fall in one of the six categories (Figure\ref{subfig:ddmOverview}):
        \begin{itemize}
            \item Supervised linear models: These are linear models that are trained using labeled data, where the target variable is known. Examples include linear regression and logistic regression. Linear models are simple and efficient and can be used for tasks such as prediction, classification, and feature selection.
            \item Unsupervised linear models: These are linear models that are trained using unlabeled data where the target variable is unknown. Examples include principal component analysis (PCA) and linear discriminant analysis (LDA). Unsupervised linear models can be used for tasks such as dimensionality reduction, feature extraction, and clustering.
            \item Supervised non-linear models: These are non-linear models that are trained using labeled data. Examples include decision trees, support vector machines (SVMs), and random forests. Non-linear models are more flexible than linear models and can capture complex relationships between the features and the target variable.
            \item Unsupervised non-linear models: These are non-linear models that are trained using unlabeled data. Examples include self-organizing maps (SOMs) and autoencoders. Unsupervised non-linear models can be used for tasks such as clustering, anomaly detection, and data compression.
            \item Supervised deep learning: These are deep learning models that are trained using labeled data, typically with large amounts of data and complex architectures. Examples include convolutional neural networks (CNNs) for image recognition, recurrent neural networks (RNNs) for natural language processing, and fully connected deep neural networks for regression and classification tasks. Deep learning models can achieve state-of-the-art performance on many tasks and are highly flexible.
            \item Unsupervised deep learning: These are deep learning models that are trained using unlabeled data, often with large amounts of data and complex architectures. Examples include generative adversarial networks (GANs) for image synthesis and unsupervised feature learning, and autoencoders for data compression and anomaly detection. Unsupervised deep learning models can be used for tasks such as unsupervised feature learning, anomaly detection, and generative modeling.
        \end{itemize}
        
        \paragraph{DDM in Standalone Digital Twins}
        The Standalone DT does not have a data stream from a physical asset but can be useful in the design phase before the physical asset is built. In the wind energy industry, the performance of wind turbines depends heavily on their position and design~\cite{Asadi2021wfs, 2015htd}. The placement of a wind turbine requires consideration of various factors, such as wind conditions, terrain, connectivity, and risk~\cite{Asadi2021wfs}. Wind conditions, such as wind density, wind speed shear, turbulence intensity, and directional shear, play a crucial role in the design and operation~\cite{Wagner2010sos}. However, historical data may not always be available and long-term measurements may be infeasible. In such cases, machine learning techniques can be used to estimate wind conditions. For example, in \cite{Velo2014wse}, multilayer perceptrons are used to estimate the annual average wind speed at sites with complex terrain based on site-specific short-term data and data from nearby stations. Similarly, \cite{Ulkat2018pom} predicts the mean monthly wind speed based on geographical and atmospheric data using neural networks. In \cite{Asadi2021nnb}, a hybrid approach that combines multilayer perceptrons with multi-criteria decision-making is proposed for improved and adaptable wind farm siting. Additionally, genetic algorithms have been used for farm layout optimization in numerous studies, such as \cite{Yang2018oow, Grady2005pow, Emami2010nao}, while a dynastic optimization algorithm is applied in \cite{Massan2020anm} and yields results comparable to genetic algorithms. Particle swarm optimization is used to improve wind farm layout in \cite{Pillai2018owf}, while a cyber swarm algorithm is employed for micrositing in \cite{Yin2016apd}. As a user-case example, GE has used a Standalone DT to select the optimal turbine configuration for a given site based on a modular turbine design~\cite{2015htd}. The Standalone DT is therefore useful for various wind energy-related applications, particularly in the design phase, where the optimization of turbine position and design can have a significant impact on the overall performance of a wind farm.
        
        \paragraph{DDM in Descriptive DT}
        The Descriptive DT is connected to the physical asset through a (real-time) data stream. As such, the data-driven methods mentioned in Section~\ref{ssec:data_generation} (e.g. Principal Component Analysis, Generative Adversarial Networks, Autoencoders) are of interest for realizing a Descriptive DT.
        Sophisticated Descriptive DT cannot just rely on sensor data, but have to be able to estimate quantities between sensors and derived quantities through models.
        Physics-based high-fidelity models require significantly larger computational resources and are typically not able to run in real-time. This is especially true for fluid dynamics simulations with turbulence.
        It is important to note that the Descriptive DT does not use only simulations. Instead, simulations have to be applied to enhance the measured data. An example of such a procedure is the resolution enhancement of images through, e.g. Generative Adversarial Networks.
        A significant drawback of data-driven models is the lack of generalizability, i.e. adaption to unanticipated situations. This is further addressed in Section~\ref{sssec:ham}.
        
        \paragraph{DDM in Diagnostic DT} 
        Condition Monitoring and Condition based maintenance are core features of the Diagnostic DT. This requires the definition of conditions, i.e., parameter thresholds, at which the DT triggers an alarm. In many cases, data-driven models are employed to capture complex physical behavior.
        In the context of wind power,~\cite{Kruger2013add} apply PCA and calculate the test statistic and Hotelling's $T^2$. If they become too large, the reconstruction-based contributions are calculated to identify the responsible parameters. Alternatively, they define a normal operating class with Fisher Discriminant Analysis and monitor deviations through test statistics.
        Furthermore, neural-network-based techniques like Autoencoders and specifically Denoising Autoencoders (DAE) are considered for fault detection. Here, the assumption is that they are able to capture complex connections between inputs that simpler models cannot identify. In AE, the encoder compresses the input information into a vector (layer) smaller than the input and output vector. If the input is not covered by the training set, i.e. the turbine data under usual, fault-free conditions, there is a risk that the decoder is not able to reconstruct the original data set. By defining a reconstruction error, the amount of difference between input data and fault-free data is quantified. Should this reconstruction error become too high, a potential fault is detected. A DAE purposely corrupts part of the input for higher accuracy.
        This approach is used in e.g.~\cite{Wang2018wtb} and~\cite{Jiang2018wtf} on SCADA data.~\cite{Chen2021ats} presents a Generative Adversarial Network (GAN) for self-setting reconstruction error thresholds. In~\cite{Wang2022ada}, GANs and Siamese encoders (the data is encoded a second time and the coded data is compared for the reconstruction error) are used with a transfer layer to reduce the impact of ambiguous training data. 
        
        \paragraph{DDM in Predictive DT}
        The predictive DT performs forecasts by extrapolating current and recorded data in time. Variables of interest could be wind speed, direction, or turbulence, produced power, remaining useful lifetime (RUL) of components, or even the impact of climate change on the farm. As such time frames can range from seconds to decades. In either case, forecasts are frequently performed with data-driven models based on a given time series. 
        Some simpler data-driven models are, e.g., auto-regressive (AR) and moving-average (MA) models, or a combination of those with exogenous variables (X), seasonal effects (S), trend removal through differentiation and integration (I), or prediction of multiple series simultaneously through vector inputs (V).  
        The ARIMA model, a combination of the above-mentioned techniques, is used in~\cite{Yunus2016abf} for wind speed modeling.~\cite{TenaGarcia2019fod} uses a SARIMA and an NN-based model for daily wind power forecasting for each next day over a year, and mentions that a pure SARIMA model was not sufficient and outperformed by the NN-based model. 
        Again, NN-based models have been used to include complex connections between data points. ES-RNN~\cite{Smyl2020ahm}, which combined statistical methods with NNs, won the 4th Makridakis (forecasting) competition \cite{Makridakis2018tmc}, and in the 5th competition, all top-performing models were pure NN models~\cite{Makridakis2022tmu}.
        In~\cite{Manero2019dit}, Multilayer Perceptrons, Convolutional Neural Networks (CNN), Recurrent Neural Networks (RNN), and k-NN are compared for wind speed forecasts.
        
        \paragraph{DDM in Prescriptive DT}
        The Prescriptive DT provides recommendations to human operators. The DT does not automatically apply the recommendation. Therefore, recommendations must be sufficiently long-term oriented so that a human can react to them and use the recommendations to make decisions. This could include recommendations for when to perform maintenance while considering the cost of a maintenance trip against the likelihood and severity of a failure, or parameter optimization during quasi-steady states.
        These types of recommendation require control algorithms. Many control algorithms are based on or supported by physical models. An advantage of data-driven controllers is that they do not depend on a model that can potentially be flawed. However, they typically take longer to converge. 
        
        \paragraph{DDM in Autonomous DT}
        The Autonomous DT resembles the Prescriptive DT, but actions are performed immediately, without human interference in the decision process. This allows making decisions also on time scales much shorter and more frequently than possible if waiting for human approval. Data-driven control algorithms have been applied to, e.g., wind farm wake steering in simulations. Examples include gradient descent~\cite{Johnson2012aoe} and game theory~\cite{Marden2013amf}.
        
        While at earlier DT levels, there was always a human confirmation required before the DT affects the physical asset, this is removed in an autonomous DT. Therefore, every decision must be absolutely safe and reliable. The black-box nature of data-driven models makes it very difficult to guarantee the required reliability. Indeed, to the best of our knowledge, there have been no proof-of-concept tests with data-driven farm-level controllers on real wind farms. However, a reinforcement learning NN has recently been applied to a real hydrogen fusion reactor for plasma balancing, which proves that data-driven controllers can be operated even in safety-critical environments. Note that the reinforcement learning algorithm had previously undergone extensive training on a high-quality physics model~\cite{Degrave2022mco}.

        Some of the advantages of these models are online learning capability, computational efficiency for inference, and accuracy even for very challenging problems as far as the training, validation, and test data are prepared properly. However, due to their data-hungry and black-box nature, poor generalizability, inherent bias, and lack of robust theory for the analysis of model stability, their acceptability in high-stakes applications like DTs and autonomous systems is fairly limited. In fact, the numerous vulnerabilities of Deep Neural Networks (DNNs) have been exposed beyond doubt in several recent works~\cite{Akhtar2018toa,Ren2020aaa,Xu2020aaa}.

        \subsubsection{Hybrid Analysis and Modeling}
        \label{sssec:ham}

        \begin{figure}[!htb]
        \begin{subfigure}{\linewidth}
        \centering
    \includegraphics[width=\linewidth]{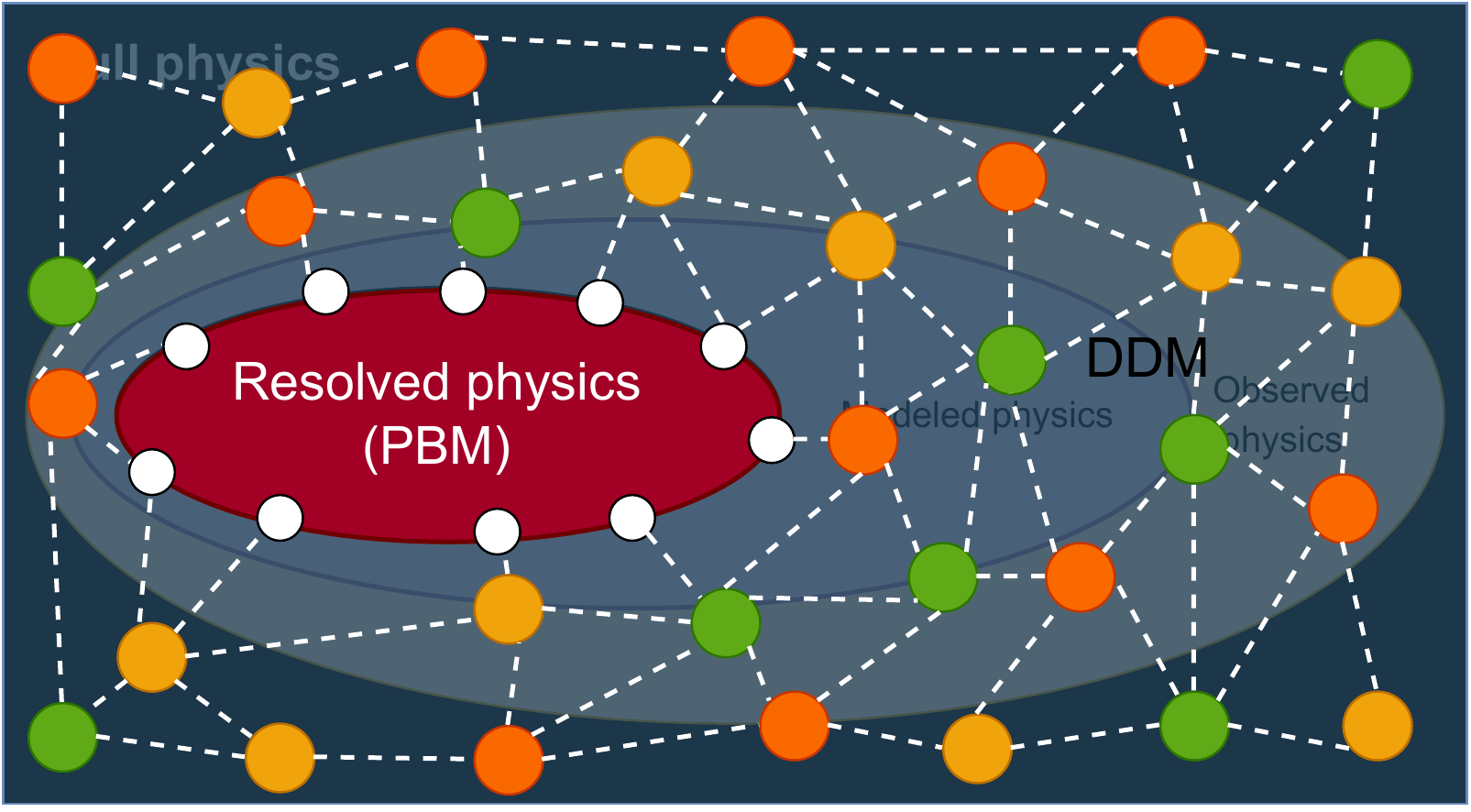}
        \caption{HAM maximizes the utilization of PBM} \vspace{0.5cm}
        \label{subfig:hamschematic}
        \end{subfigure}
        \begin{subfigure}{\linewidth}
        \centering
        \includegraphics[width=\linewidth]{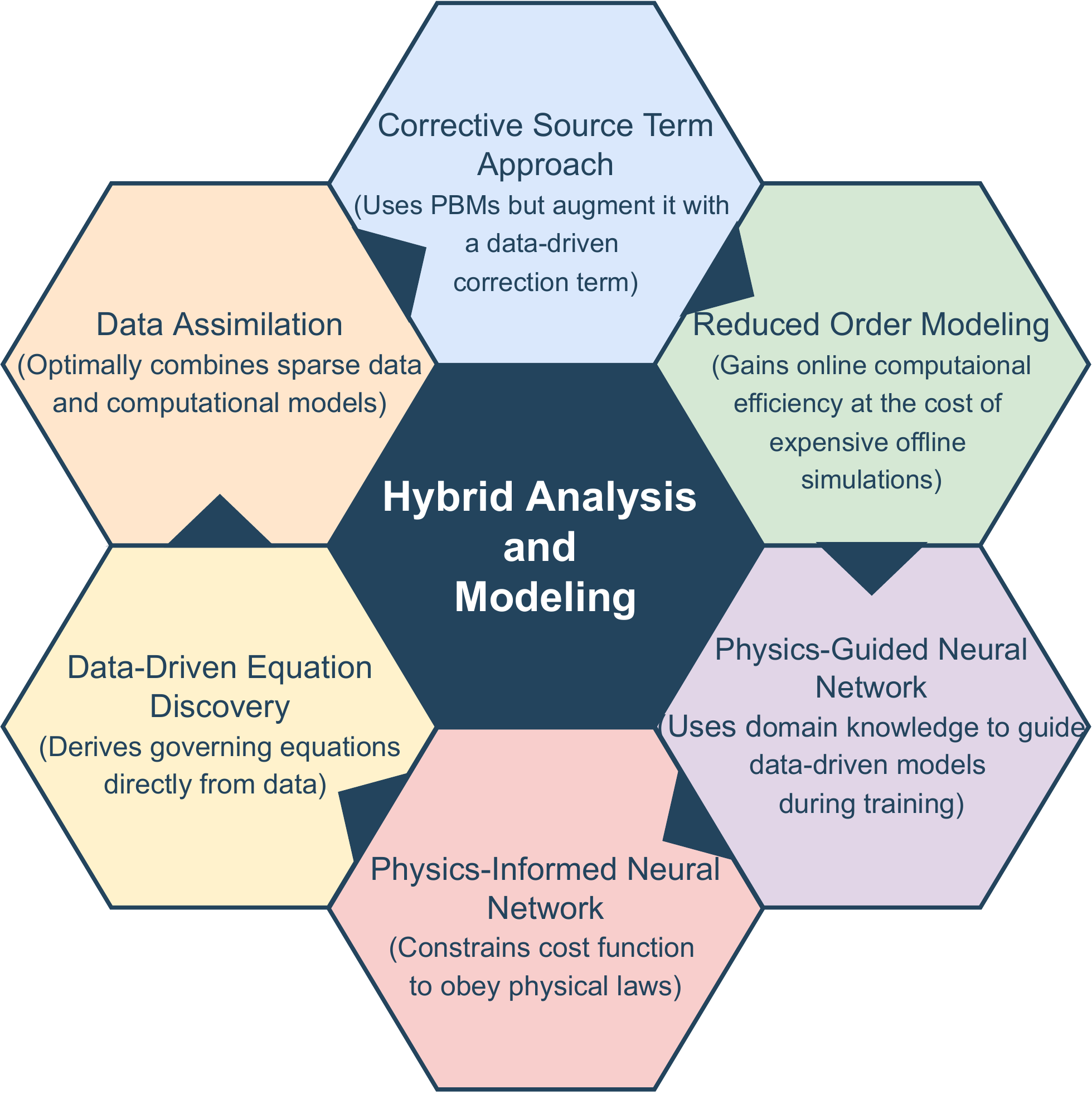}
        \caption{Six broad categories of HAM} 
        \label{subfig:hamoverview}
        \end{subfigure}
        \caption{Hybrid analysis and modeling}
        \label{fig:HAM}
        \end{figure}
        
        \begin{table*}
        \caption{\label{tab:PBMvsDDM} Physics-based modeling vs data-driven modeling}
        \begin{tabular}{p{5.0cm} p{5.7cm} p{5.7cm}}
        \hline
        Characteristic & Physics-based modeling & Data-driven modeling\\
        \hline
        Reliability & $+$ Based on well-known physics and reasoning. & $-$ Many advanced methods work like black boxes, making it difficult to fully understand their predictions. \\
        \hline
        Stability & $+-$ Nonlinear models can be sensitive and susceptible to numerical instability due to a range of reasons (boundary conditions, initial conditions, uncertainties in the input parameters), but well-established physics provides a basis for stable models. & $-$ Models can be unstable between data points and sensitive to noise, especially when overfitted to the training data. \\
        \hline
        Dealing with unknown physics & $-$ Can only account for known and implemented physics, limiting their scope of use. & $+$ Can predict from data even if the underlying physics are unknown, expanding the range of applications. \\
        \hline
        Ever-evolving (Adaptability to unexpected situations) & $-$ Only works under conditions that were accounted for in the design, making it difficult to adapt to new or unforeseen situations. & $+$ Advanced methods such as reinforcement learning can adapt to changes and learn from new data. \\
        \hline
        Generalizability to similar problems & $+$ Based on physical principles with application to a wide range of problems, as long as the physics is well understood. & $-$ Dependency on training data limits their usage to specific problems and applications. \\
        \hline
        Susceptibility to bias & $+$ Bias limited to potential bias from approximation or model assumptions. & $-$ Bias in data directly translates into model predictions, potentially leading to unfair and inaccurate outcomes. \\
        \hline
        Interpretability & $+$ The model output can be traced back to the input through physical laws and assumptions, providing a clear understanding of the model behavior. & $-$ Black-box nature of many advanced methods makes it challenging to understand how the model arrived at its predictions. \\
        \hline
        Computational cost online & $-$ Complex problems require significant computational resources and may not be executable in real-time, limiting their use in time-critical applications. & $+$ Most methods work significantly faster than real-time even for very complex problems, enabling real-time predictions. \\
        \hline
        Computational cost offline & $+$ No offline computation is required except for optional validation, making them easy to deploy. & $-$ Tuning of model parameters can take significant time and computational resources, requiring specialized knowledge and expertise. \\
        \hline
        Accuracy & $+-$ Accuracy is based on the level of detail in physics models and their ability to capture the relevant phenomena but can be limited by model assumptions and simplifications. & $+-$ Accuracy is based on the quality of data, model design, and sometimes even chance, but can be improved through the use of advanced algorithms and techniques. \\
        \hline
        Uncertainty prediction & $+$ Uncertainty can be bounded and estimated based on the assumptions and uncertainties in the physics models. & $+-$ Only some methods allow the estimation of uncertainties, which is critical for decision-making and risk analysis. \\
        \hline
        \end{tabular}
        \end{table*}

        At the bare minimum, to instill physical realism in a DT, one desires at least the following characteristics in any modeling approach: 
        \begin{itemize}
        \item accuracy
        \item computational efficiency
        \item trustworthiness
        \item generalizability
        \item self-evolution
        \end{itemize}
        
        A model's generalizability refers to its ability to solve a wide variety of problems without any problem-specific fine-tuning. Trustworthiness refers to the extent to which a model is explainable, while computational efficiency and accuracy refer to the model's ability to make real-time predictions that match ground truth as closely as possible. Lastly, a model is self-adapting if it can learn and evolve when new situations are encountered. 
        PBMs can achieve high fidelity, but at a computational cost that is not available in the DT context. Specifically for CFD, a detailed simulation modeling a few seconds often requires weeks or even months of computation time on high-power computing clusters. Furthermore, PBMs are not self-evolving but fixed to the pre-programmed models. Data-driven models are typically much faster, but their complexity and black-box nature lack the trustworthiness that is often required for industrial applications, especially in safety-critical situations like fault detection. Furthermore, data-driven models are typically applied to a very specific task and, once trained, are not able to generalize to new scenarios. Finally, while data-driven models like Reinforcement Learning are self-evolving, they are based on a trial-and-error approach that cannot be allowed in a real environment. A brief comparison of the PBM and DDM is given in Table~\ref{tab:challengesenablers}. It can be concluded that neither of the modeling approaches is an ideal candidate for usage in a DT context.

        Fortunately, a new paradigm in modeling called Hybrid Analysis and Modeling (HAM), (Figure~\ref{subfig:hamschematic})-- which combines the generalizability, interpretability, robust foundation, and understanding of PBM with the accuracy, computational efficiency, and automatic pattern-identification capabilities of advanced DDM, in particular DNNs -- is emerging. Grey box models and hybrid semi-parametric models fall into this category. While certain HAM approaches have been investigated for decades, their popularity and impact have only increased during the last few years as DDMs are becoming increasingly successful, but also more complex and less interpretable \cite{Thompson1994mcp, Venkatasubramanian2019tpo, VonStosch2014hsp}.
        In their recent surveys, authors in~\cite{Willard2022isk}, and~\cite{San2021haa} provide comprehensive overviews of techniques to integrate DDM with PBM (see also \cite{Karpatne2017tgd,Willcox2021tioa,Ihme2022cml}). Most hybridization techniques, as shown in Figure~\ref{subfig:hamoverview}, fall into one of the following categories:
            \paragraph{Corrective source term approach}
            Corrective source term approach (CoSTA) is a method proposed in~\cite{Blakseth2022dnn} that explicitly addresses the problem of unknown physics. This is done by augmenting the governing equations of a PBM describing partial physics with a DNN-generated corrective source term that takes into account the remaining unknown/ignored physics. One added benefit of the CoSTA approach is that the physical laws can be used to keep a sanity check on the predictions of the DNN used, i.e. checking conservation laws. A similar approach has also been used to model unresolved physics in turbulent flows~\cite{Maulik2019smf,Pawar2020apa}. CoSTA can enable coarse-scale simulations without any loss of accuracy since the ignored subgrid scales are compensated through a data-driven source term. The coarse-scale simulations can be useful at almost all the capability levels. The approach can be used for optimizing wind farm layout through accurate wake modeling, filling in the coarse spatiotemporal resolution of measured data, diagnosing anomalies through the analysis of the source term, and making real-time predictions about the future state of the asset. However, even these approaches assume a specific structure for at least the known part of the equation.
            \paragraph{Data assimilation}
            Data assimilation is one of the strongly rooted methodologies that combines dynamical models with observational data and has a long history of decades in numerical weather predictions \cite{Lewis2006dda,Lakshmivarahan2017fec}.
            In DDM paradigms, there is always the notion of offline training and online deployment. A fundamental challenge in these approaches is to address and tackle poor generalization to distributional shifts in the data (e.g., generalizing beyond training conditions). Therefore, the applicability of the DDM models is usually limited by the training algorithm and the training data sets. However, a DT with various DDM components should self-adapt to the new condition as it evolves. Accordingly, data assimilation algorithms might provide this capability by making use of available streams of sensor measurements from the physical system. These data assimilation algorithms can be also exploited for the parameterization of better models to enable improved corrections to the DDM dynamics \cite{Pawar2021anh,Pawar2021dae,Ahmed2020fsa}. Moreover, as was discussed earlier in Section~\ref{ssec:data_generation}, such techniques can be used to optimize the experimental configuration and sensor placements to decrease the costs of data collection, and improve the quality of the inference algorithms \cite{Deng2021dnn,Uilhoorn2022aaf}. Recent discussions on model-data fusion and integration of DDM and data assimilation approaches can be found in \cite{Gettelman2022tfo,Geer2021les,Brajard2021cda,Buizza2022dli}. Data assimilation can be extremely useful for predictive DT.
        
            \paragraph{Data-driven physics discovery}
            One of the challenges which can really jeopardize the functioning of a digital twin are those phenomena about which there is no complete understanding, and hence the equation cannot be written down from first principle to model. Although some of the other methods try to compensate for this using a data-driven approach, they do not help in developing new understanding and discovering the new physics. To this end, sparse regression based on $l_1$ regularization and symbolic regression based on gene expression programming have been shown to be very effective in discovering hidden or partially known physics directly from data. Notable work using this approach can be found in~\cite{Brunton2016dge},~\cite{Champion2019ddd}, and~\cite{Vaddireddy2020fea}. Provided that enough high-resolution LIDAR data are available, the approach can be used to derive better mathematical models of wakes directly from the data. Another application could be to use infrared thermography data to derive equations, which can be later used to detect anomalies in gearboxes. However, one of the limitations of this class of methods is that, in the case of sparse regression, additional features are required to be handcrafted, while in the case of symbolic regression, the resulting models can often be unstable and prone to overfitting.

            \paragraph{Physics guided neural network}
            One of the active research thrusts is to leverage methodologies for the combination of physics-based and neural network models, a rapidly emerging field that came to be known as physics-guided NN (PGNN) or physics-guided ML (PGML). To this end, we have recently introduced a PGML framework in which information from simplified PBMs is incorporated within neural network architectures to improve the generalizability of data-driven models~\cite{Pawar2021pgm, Robinson2022pgn}. The central idea in the PGNN framework is to embed the knowledge from simplified theories directly into an intermediate layer of the neural network. The approach can be used to fuse data and different types of models~\cite{Pawar2022mfi,Pawar2021mfw}. The knowledge from the simplified theories aids in ensuring that we learn only the knowledge required to compensate for the deficiencies of these theories instead of learning everything from scratch. Also, owing to the fact that the simplified theories are still based on the laws of nature they are more generalizable compared to any data-driven approach and hence they should be exploited to the extent possible. For example, the prediction of flow around an airfoil is a high-dimensional and nonlinear problem that can be solved using high-fidelity methods like computational fluid dynamics (CFD). PGNN/ML can be used for fast modeling of the aerodynamic characteristic of turbines that can be used in real-time control systems. More recently, the release of the Theseus library \cite{Pineda2023tal} is facilitating the research in this direction.   
            
            \paragraph{Physics informed neural network}
            By incorporating a PBM in the objective function, DDMs can be biased toward known physical laws during training. A prominent recent work by~\cite{Raissi2019pin} is the physics-informed neural network (PINN), where a NN is used to represent the solution to a PDE, and deviations from the equation at a sample of points are penalized by an additional loss term. PINNs can be used to solve problems such as heat transfer, as was done by~\cite{Zobeiry2021api} for parts in a manufacturing process. The PINN approach has also been extended by~\cite{Arnold2021ssm} to allow for control in a state-space setting. In related work, the researchers of~\cite{Shen2021api} create a model for classifying bearing health by training a NN on physics-based features and regularizing the model using the output of a physics-based threshold model. A problem with these approaches is that they require precise knowledge of the loss term. The regularization can also pose a challenge during the training process because of the increase in the complexity of the cost function, especially if computing the regularization term requires the evaluation of a complex model. The readers are referred to~\cite{Cuomo2022sml} and references therein for a recent state-of-the-art discussion on where PINNs are and what might come next.
                       
            \paragraph{Reduced order modeling}
            Within the model order reduction approach, reduced-order modeling (ROM) has been very popular~\cite{Quarteroni2014rom}.  In ROM, full-order models (FOM) are projected onto a reduced-dimensional space based, e.g., on the proper orthogonal decomposition of the FOM simulation results (snapshots)~(\cite{Chinesta2011asr,Fonn2019fdc,Quarteroni2014rom,Brunton2020mlf,Taira2020mao}). Provided that the information in the FOM results can be retained with a considerably reduced dimension (such that the truncation of the dimension does not lead to significant error), one can achieve a stable ROM with several order of magnitude speedup. However, these models have two limitations in the context of realistic problems. Firstly, the truncation of the dimensions tends to destabilize the model. Secondly, these models are often intrusive in the sense that both the original equations and the data are required to build the ROM. To address the first issue, eddy-viscosity-based ROMs~\cite{Osth2014otn,san2014proper,Hijazi2020ddp,Tsiolakis2021rom} or semi-intrusive ROMs~\cite{Pawar2021dae,Ahmed2019men,Pawar2020ddr,Pawar2020aet,Wang2020rnn,Ivagnes2023hdd} where the effect of the ignored modes are corrected using a data-driven approach have been proposed. Moreover, fully non-intrusive ROMs \cite{Xiao2015nir,Yu2019nir,Rahman2019nro,Siddiqui2019fvh,Ahmed2021npo,Benner2020oif} are gaining popularity because of their ability to address the second issue. For a comprehensive review of the ROM methodology, the readers are referred to \cite{Ahmed2021ocf}. 
    
    \subsection{Industrial acceptance}
    \label{ssec:Industrialacceptance}
    DTs are specifically targeted at industrial applications. This holds true, especially for DTs related to wind energy. Operators, manufacturers, and consulting companies have to collaborate on DTs to extract the full benefits. As such, technology must be researched and developed to the point where the industry can be reasonably expected to continue industrial research and apply it to its assets.
    \subsubsection{Technology Readiness Levels}
    The Technology Readiness Level (TRL) scale provides a measurement for the maturity of a technology or innovation from initial basic research to competitive usage within the industry. First used by NASA, it is now widely applied, for example, in the EU Horizon 2020 program and the NorthWind project~\cite{NationalAeronauticsandSpaceAdministration2012trl,EuropeanCommission.DirectorateGeneralforResearchandInnovation.2017trl,NorthWindnrc}.
    The TRL scale consists of 9 levels~\cite{EuropeanCommission.DirectorateGeneralforResearchandInnovation.2017trl} as 1: Basic principles observed, 2: Technology concept formulated, 3: Experimental proof of concept, 4: Technology validated in lab, 5: Technology validated in relevant environment, 6:Technology demonstrated in relevant environment, 7: System prototype demonstration in operational environment, 8: System complete and qualified, 9: Actual system proven in operational environment.
    
    The TRL describes how close a technology is to industrial application. In the context of DTs, it is orthogonal to the capability level scale. Each capability level can be treated as a separate technology. One might argue that research should first perform research on an Autonomous (level 5) DT and only then elevate its TRL to industrial maturity. This approach would significantly delay the deployment of DTs within the industry. In fact, the capability levels are defined specifically to allow unlocking significant value at each level, as confirmed by the industry survey. Additionally, each level builds on components from all previous levels. By developing each capability level to a point where the industry can adapt the DT, value is generated much sooner. In addition, the industrial infrastructure adapts much earlier to DTs. This includes sensor installations, data acquisition, acclimatization of staff, and establishment of a workforce specialized for DT development and operation. All these factors are relevant from the earlier levels onward and become more and more important as the DTs become more capable. Progressing along the capability levels, therefore, allows a gradual adaptation of DTs in the industry, thereby also lowering the acceptance threshold gradually. Recommendations and control of Prescriptive and Autonomous DTs naturally find more acceptance if the underlying data acquisition, modeling, analysis, and forecasting techniques are already well-established within the industry. 
    It is noteworthy that the TRL is specifically suited for technology development and that there are more complex (higher-dimensional) scales including more factors and stages. An example is the Balanced Readiness Level assessment proposed in \cite{Vik2021brl}. It combines the development-oriented TRL with readiness level assessments from legalization, commodification, (public) acceptance, and compatibility with existing technologies and practices. 

    \subsubsection{Human machine interface}
    Another important factor for industrial and public acceptance of DTs is the interface. An intuitive interface is essential to navigate the large variety of data collected in a DT. Furthermore, it is important to address the needs of all potential users when designing the interface.
    \begin{figure}
    \centering
    \includegraphics[width=\linewidth]{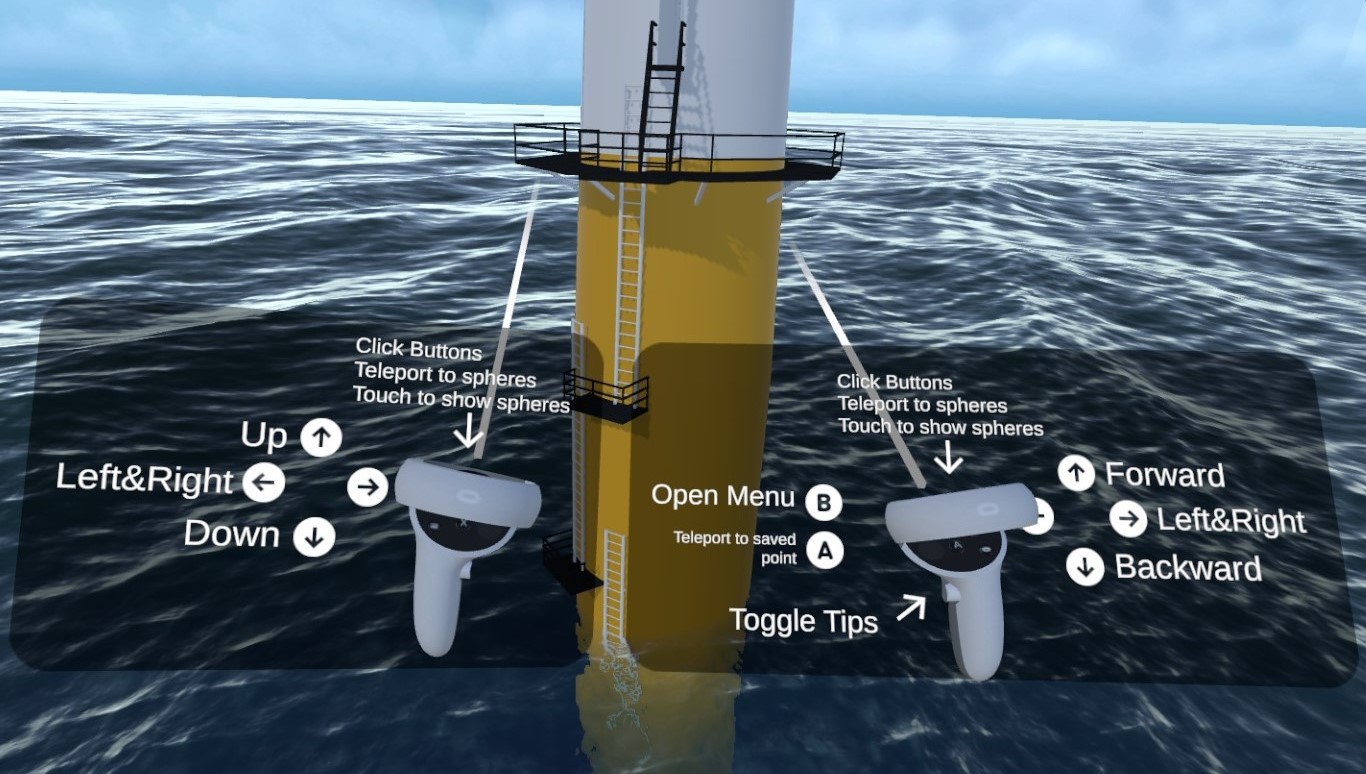}
    \caption{VR controllers with tooltips.}
    \label{fig:controllers}
    \end{figure}
    The advantages of a 3D model can be tremendous for data visualization. Simply having the relevant data values shown on the individual components may be beneficial. Advanced examples include heat map overlays for visualization of temperature, stress, component wear or vibrations, or color encoding of critical components in a turbine. This is especially advantageous for stakeholders without knowledge of the technical details.
    Immersive media, specifically enhanced reality (XR) can be used to improve human-machine interaction. Instead of presenting the 3D interface on a 2D screen, virtual reality (VR) can be utilized. In VR, the user experiences a completely virtual environment. In augmented reality (AR) and mixed reality (MR), virtual information overlays the physical environment. This enables visualization of the data on the real components and is a promising tool to increase the efficiency of maintenance work. Currently, however, AR technology is still rather expensive. Additionally, computational resources have to be considered for such a mobile AR lab DT. However, there is no doubt that giving a hands-on experience of the DT is the best way to communicate the untapped potential of the DT technology. 
    
    \subsubsection{Relevant User-Cases}
    \label{sssec:usercases}
    \begin{figure*}[!htb]
       \begin{subfigure}{\linewidth}
         \centering
         \includegraphics[width=\linewidth]{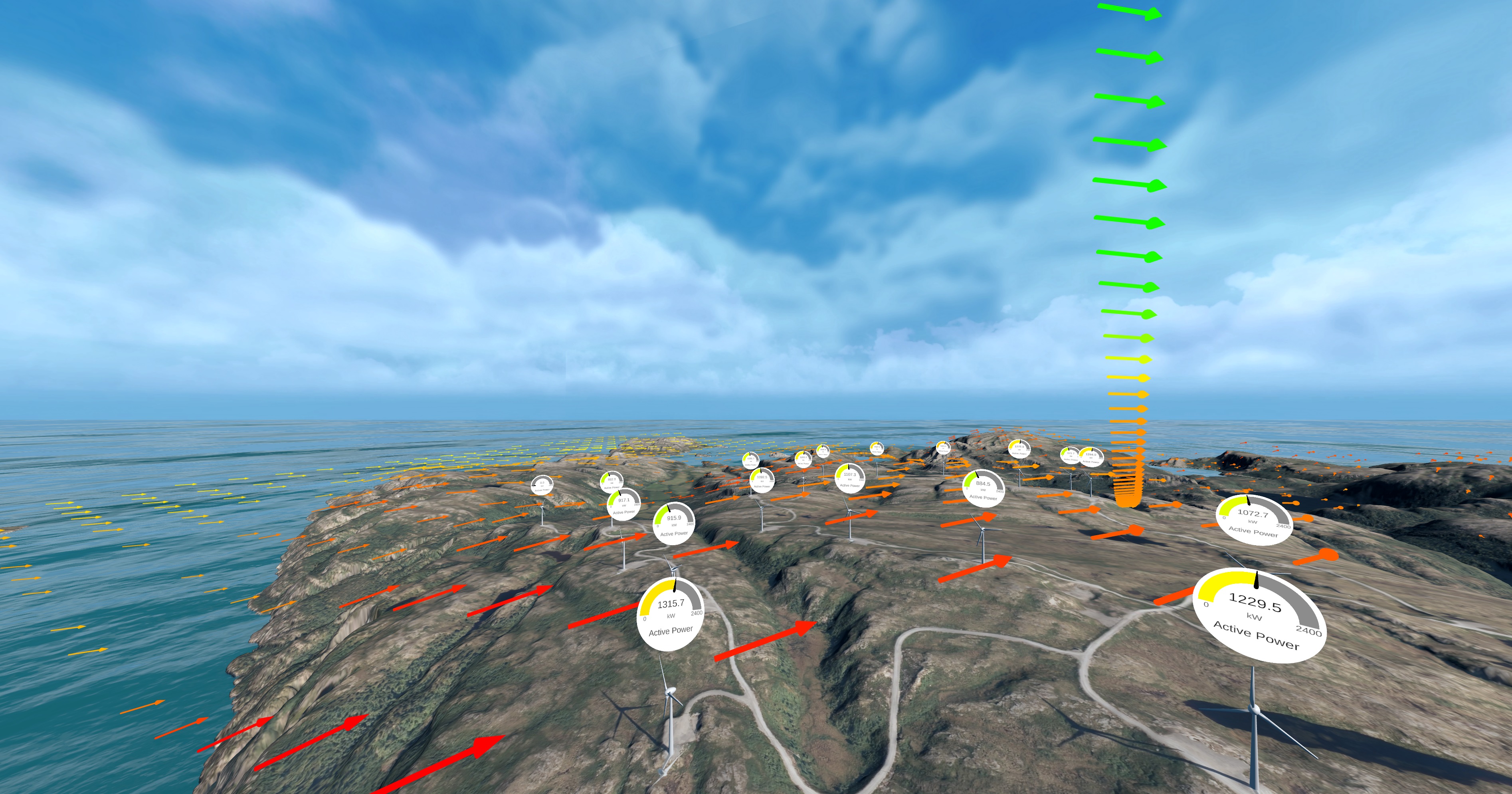}
         \caption{Descriptive DT of the Bessaker wind farm showing terrain, realtime data, and numerically predicted wind field}\vspace{0.5cm}
         \label{subfig:BessakerVR}
       \end{subfigure} 
       \begin{subfigure}{\linewidth}
         \centering
         \includegraphics[width=\linewidth]{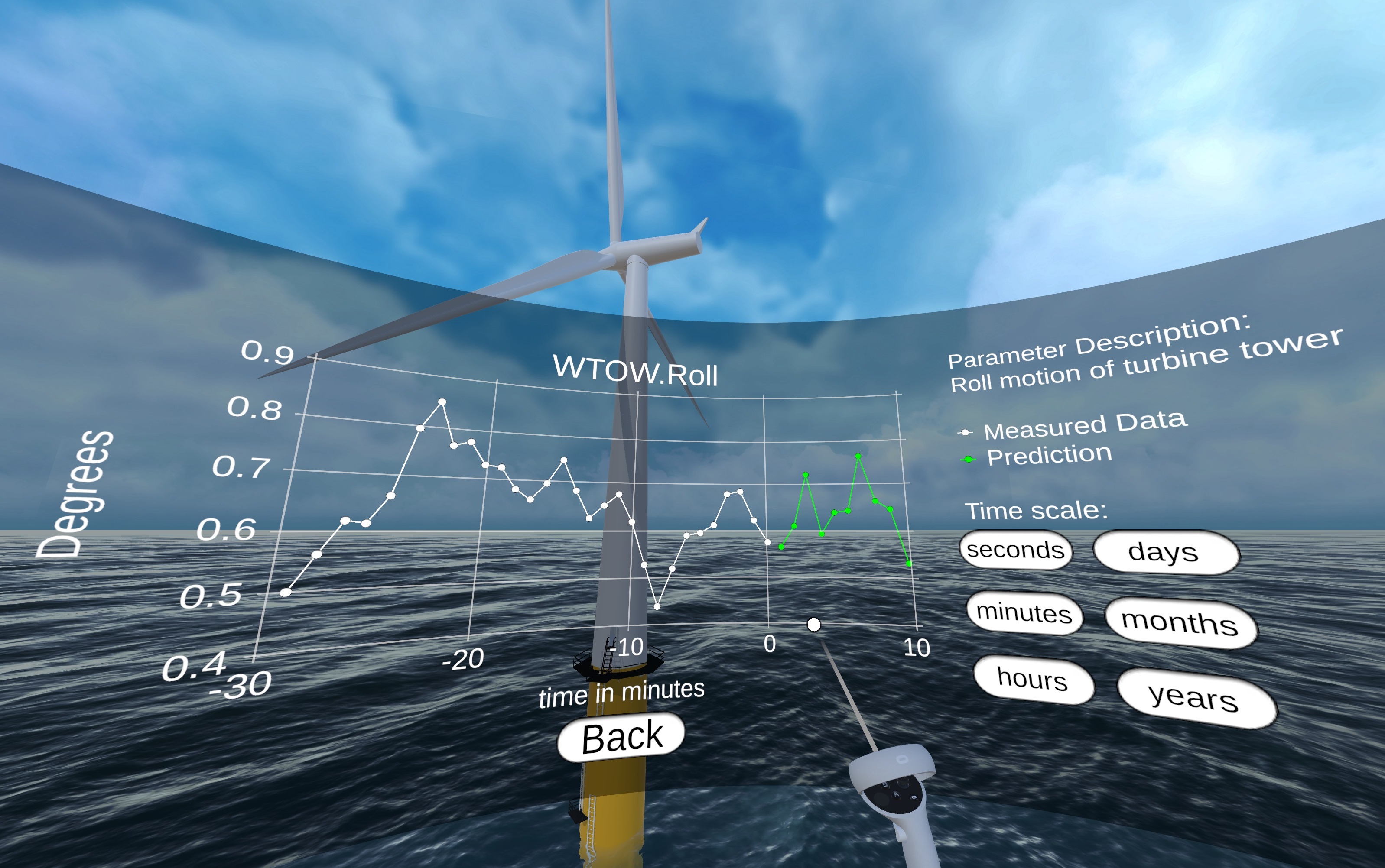}
         \caption{Descriptive and predictive DT of the Zefyros wind turbine}
         \label{subfig:ZefyrosVR}
       \end{subfigure}
        \caption{Demonstration of an onshore and offshore use case in virtual reality}
        \label{fig:usecase}
    \end{figure*}
    
    An essential part of the TRL scale is the test and validation of the technology in a real environment. Industrial acceptance will not happen without proof of DTs generating value from real assets. Once there is proof that the technology brings profit, it will be much easier for industries to justify funding for in-house DT development. As such, relevant user cases are of great importance for the transition from academic to industrial application. The realization of such user cases is, however, rather difficult. Already the Standalone DT benefits from historic site or asset data, design data, or even a CAD-model. From level one onward, additional (pseudo-) real-time data are required. Both design and operational data are heavily restricted for academia, as this data counts as proprietary information. For level 5, it will become even more difficult to provide meaningful user cases, as the autonomous DT includes a feedback loop, which essentially requires farm control. Significant time has to be spent on verifying the autonomous DT in simulations before it will be used in the real world. 
    It is essential for both academia and industry to closely collaborate on user cases through data and model sharing to establish DTs as a new standard technique. As an example, in the NorthWind project, two use cases are chosen, one for onshore (Figure~\ref{subfig:BessakerVR}) and one for offshore (Figure~\ref{subfig:ZefyrosVR}). A user can interact with the DT using the interface and controllers as shown in Figure~\ref{fig:controllers}. 
    The 3D models of the turbines and the environment have been created completely from openly accessible information, but the data measured at the turbine are confidential (the figures are created with mock data). More information on the user cases can be found in \cite{Stadtmann2023sda, Stadtmann2023doa}. Having a realistic use case in the project has already started showing the benefits as the development of various enabling technologies keeps the end use in view. It is hoped that these use cases will help in tighter collaboration between different stakeholders. 
      
\section{Conclusion and recommendations}
\label{sec:conclusion} 
\begin{table*}[ht]
\caption{Mapping between common challenges and enabling technologies}
\label{tab:challengesenablers}
\begin{tabular}{p{8.5cm} p{8.5cm}}
\hline
Challenges & Enabling Technologies\\ \hline \hline
Proprietary data and data privacy & Federated Machine Learning, blockchain  \\
Real-time communication of large amounts of data & Data compression, internet of things, 5G  \\
Varying data formats and model interfaces & Global standardization complementing existing standards \\
Data silos & Standardization, ontologies, and asset information models\\
Data quality & Improved sensor systems, Uncertainty quantification \\
Physical realism despite spatio-temporal data sparsity & Spatio-temporal interpolation and extrapolation, physics-informed machine learning \\
Immeasurable quantities and inaccessible locations & Virtual sensing, advanced sensor technologies \\
Real-time modeling and analysis & Hybrid analysis and modeling, reduced order modeling, multivariate data-driven models, real-time optimization \\
Unknown physics & Data-driven models, unsupervised learning, data assimilation, compressed sensing, symbolic regression, physics-informed machine learning \\
Dynamic models evolving with the asset & Hybrid analysis and modeling, data assimilation, reinforcement learning, adaptive control \\
Computational expensive models & Edge, fog, and cloud computing, offline trained data-driven models, hybrid modeling, reduced order modeling \\
Transparency and interpretability & Hybrid analysis and modeling, explainable artificial intelligence, interpretable machine learning, causal inference \\
Industrial acceptance & Intuitive interfaces through VR and AR, industry involvement through pilot projects, collaboration with domain experts \\
Public acceptance & Scientific outreach, public engagement, education, and training, social media communication \\
\hline
\end{tabular}
\end{table*}

\begin{figure*}[ht!]
\centering
\includegraphics[width=\linewidth]{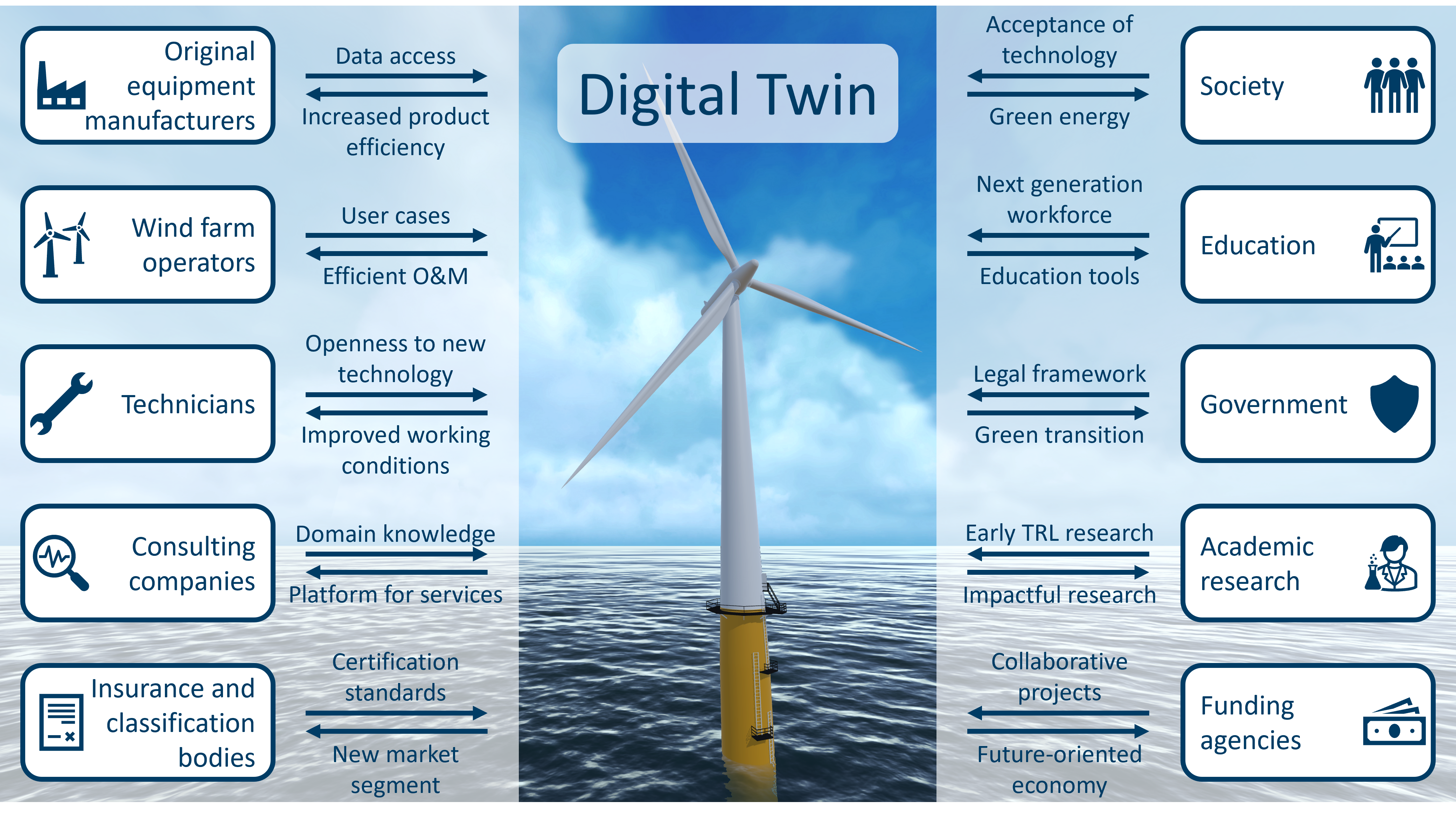}
\caption{Stakeholders and their potential contributions.}
\label{fig:stakeholders}
\end{figure*}
The paper cited a collection of definitions of digital twins (DT) and reiterated the capability level classification of DT in the context of wind energy. It then, through a rigorous literature review, identifies challenges in realizing highly capable digital twins from an industrial perspective. The main contributions of this work are:

\begin{itemize}
\item Conducting a survey to gather industry perspectives on DT technology, with a specific focus on its applications in wind energy. The survey identified several critical research challenges that need to be addressed to fully realize the benefits of DTs. These challenges were related to standards, data, models, and industrial acceptance.
\item Conducting a targeted literature survey and consulting with industry partners to identify potential solutions to the challenges identified in the survey. The results of this research are summarized in Table \ref{tab:challengesenablers}.
\item Lastly, in the following section, we provide recommendations for the roles the various stakeholder need to play for mainstream acceptance, deployment, and ultimately projection of the technology on the read assets. 
\end{itemize}

As already discussed in detail, digital twinning is an emerging technology that has the potential to revolutionize various industries, but its success will depend on the collaborative efforts of various stakeholders. Based on our technology watch, we would like to conclude our analysis by providing recommendations for each stakeholders group (Figure~\ref{fig:stakeholders}) and their most important contributions:
\begin{itemize}
    \item  \textit{Industry}: The industry sector is expected to be the biggest driver of DT technology and can contribute positively in three ways. Firstly, they can provide asset data sets for research and model building. Secondly, they can actively participate in research by sharing practical knowledge. Thirdly, they can validate the usefulness of DTs by applying the insights obtained from predictive twins into their business applications.

    \item \textit{Academia and research institutes}: Academia and research institutes are expected to play a significant role in the development of enabling technologies for both virtual and predictive twins. It is recommended that these developments are made exploitable for society at large through open-source software. Additionally, academia should take the lead in grooming a new generation of the interdisciplinary workforce by following the MAC-model, which combines application knowledge with expertise and advanced methodologies from mathematics and computer science.

    \item \textit{Government and policy makers}: They have a critical role to play in ensuring that the benefits of the new technology reach every layer of society while safeguarding ethics, privacy, and security. They should focus on framing inclusive policies and regulations that democratize the technology. For instance, they can initiate feasibility studies for utilizing DTs in their sectors and make data generated by means of public funding available for academia and industry.

    \item \textit{Funding agencies}: Funding agencies, especially those with a mission to focus on industrial innovation impact, should prioritize digital twinning as a theme for center projects. Funding of open-source enabling technology platforms should be prioritized as infrastructure funding has been scarce up to now.

    \item \textit{Society}: Finally, it is the responsibility of the society itself to be well-informed about the new technology. Starting from K12 education, society should develop new skills that will facilitate the embracement of the emerging technology. By doing so, we can ensure that the new technology is successfully integrated into our private and professional lives.
\end{itemize}   
We end this article by citing a work \cite{Elfarri2023aid} in which the authors in the context of built environment have demonstrated how a highly capable digital twin can be quickly developed if the challenges surrounding data sharing are resolved.

\section*{Acknowledgment}
This publication has been prepared as part of NorthWind (Norwegian Research Centre on Wind Energy) co-financed by the Research Council of Norway (project code 321954), industry, and research partners. Read more at www.northwindresearch.no

\bibliographystyle{AR} 
\bibliography{referencesAR,refMisc}
\newpage
\begin{wrapfigure}{l}{25mm} 
    \includegraphics[width=0.975in,height=1.3in,clip]{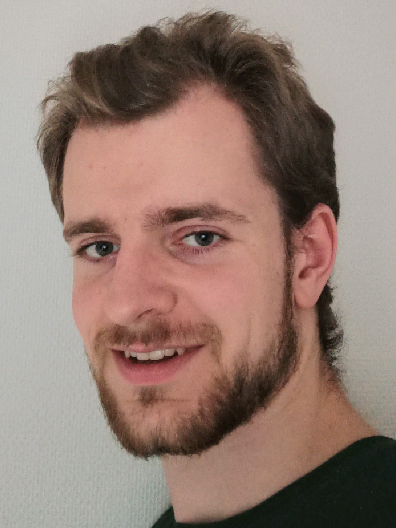}
\end{wrapfigure} \textbf{Florian Stadtmann} He is currently a Ph.D. student in the Department of Engineering Cybernetics at the Norwegian University of Science and Technology. He is conducting research on enabling technologies for Digital Twins. He received his B.Sc. in Physics and M.Sc. in Astroparticle Physics and Cosmology from RWTH Aachen University, Germany, where he focused on combining physics-based models with AI. 

\begin{wrapfigure}{l}{25mm} 
    \includegraphics[width=0.975in,height=1.3in,clip]{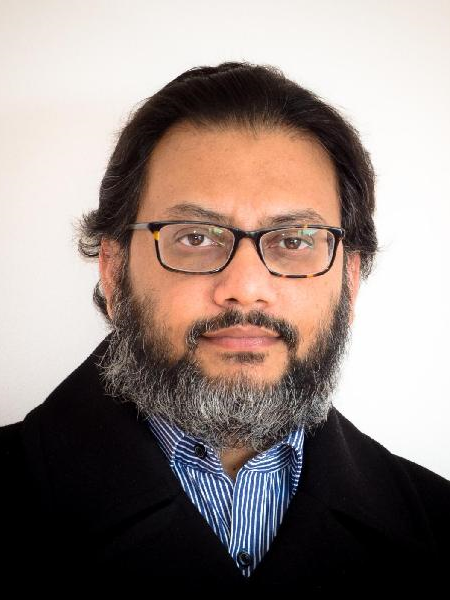}
\end{wrapfigure} \textbf{Adil Rasheed} He is a Professor in the Department of Engineering Cybernetics at the Norwegian University of Science and Technology. There, he works to advance the development of novel hybrid methods that combine big data, physics-driven modeling, and data-driven modeling in the context of real-time automation and control. In addition, he also holds a part-time Senior Scientist position in the Department of Mathematics and Cybernetics at SINTEF Digital, where he previously served as the leader of the Computational Sciences and Engineering group from 2012 to 2018. His contributions in these roles have been the development and advancement of both the Hybrid Analysis and Modeling and Big Data Cybernetics concepts. Over the course of his career, Rasheed has been the driving force behind numerous projects focused on different aspects of digital twin technology, ranging from autonomous ships to wind energy, aquaculture, drones, business processes, and indoor farming. He is currently leading the Digital Twin and Asset Management related work in the FME Northwind center.

\begin{wrapfigure}{l}{25mm} 
    \includegraphics[width=0.975in,height=1.3in,clip]{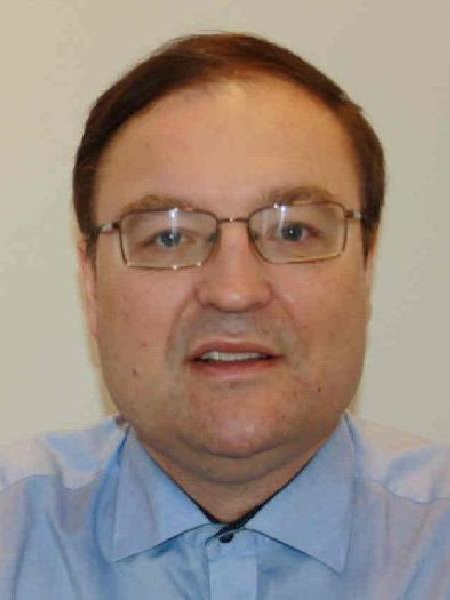}
\end{wrapfigure} \textbf{Trond Kvamsdal} He is a Professor in the Department of Mathematical Sciences, at the Norwegian University of Science and Technology, and he also holds a part time Senior Scientist position in the Department of Mathematics and Cybernetics at SINTEF Digital. His positions at NTNU are within computational mathematics, i.e. development of new theories/methods within applied mathematics and numerical analysis to make robust and efficient numerical software programs for challenging applications in science and technology. Main areas of 
research are Adaptive Finite Element Methods (AFEM), Reduced Order Modeling (ROM), and Hybrid Analysis and Modeling (HAM) to enable predictive digital twins. He is chairing the FME NorthWind Scientific Advisory Committee and is the leader of the NTNU Energy Team Wind and the NTNU IE Team Digital Twin. He received the IACM Fellow Award (International Association for Computational Mechanics) in 2010 and was elected member of the Norwegian Academy of Technological Sciences (NTVA) in 2017. 

\begin{wrapfigure}{l}{25mm} 
    \includegraphics[width=0.975in,height=1.3in,clip]{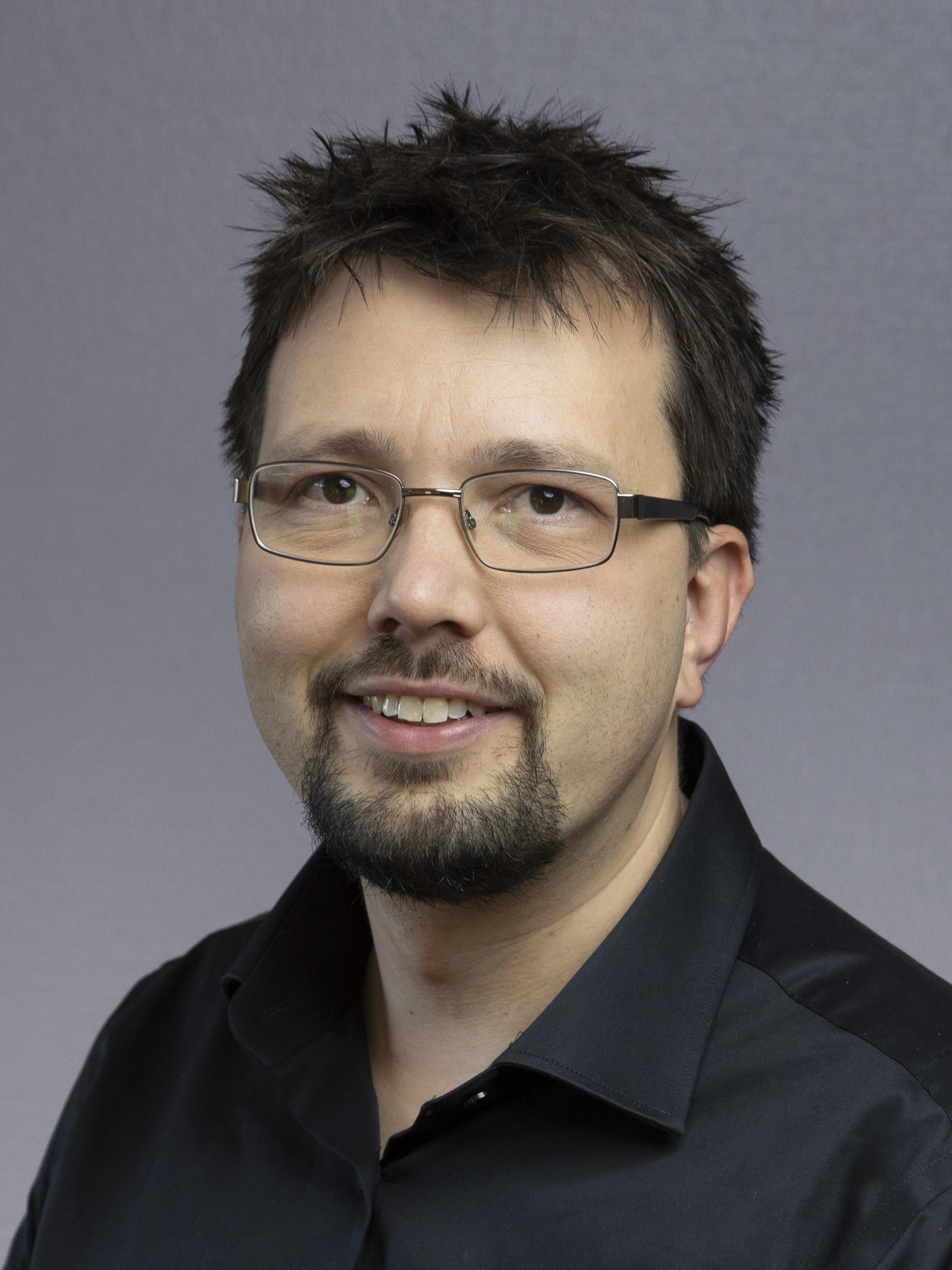}
\end{wrapfigure} \textbf{Kjetil Andr\'{e} Johannessen} He is the Research Manager of the Computational Science and Engineering (CSE) group in the department of Mathematics and Cybernetics at SINTEF Digital. He holds a PhD in Isogeometric Analysis Using Locally Refined B-splines from the Norwegian University of Science and Technology (NTNU). He has been a pioneer within Adaptive Isogeometric Finite Element Methods and his first paper on this topic is a {\it Highly Cited Paper} according to the Web of Science. He is the leader of SINTEF Digital's research area on Digital Twin and Mixed Reality and is the deputy leader for the activities on digital twin and asset management in the FME NorthWind centre.

\begin{wrapfigure}{l}{25mm} 
    \includegraphics[width=0.975in,height=1.3in,clip]{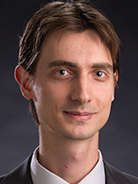}
\end{wrapfigure} \textbf{Omer San} He is an Associate Professor in the School of Mechanical and Aerospace Engineering at Oklahoma State University. He received his PhD in Engineering Mechanics from Virginia Tech. His field of study is centered upon the development, analysis and application of advanced computational methods in science and engineering with a particular emphasis on the development of hybrid physics-AI/ML methodologies in fluid dynamics across a variety of spatial and temporal scales. He has been a PI (Principal Investigator) or co-PI on research grants from the DOE, NSF, ASHRAE and NASA, and his work has been recognized through several distinguished awards and nominations including the DOE Early Career Research Program Award in Applied Mathematics to develop scientific machine learning algorithms for multiscale closure modeling of turbulent flows.

\begin{wrapfigure}{l}{25mm} 
    \includegraphics[width=0.975in,height=1.3in,clip]{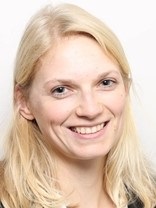}
\end{wrapfigure} \textbf{Konstanze Kölle} She works currently as Research Scientist at SINTEF Energy Research. She holds a PhD in Engineering Cybernetics from the Norwegian University of Science and Technology (NTNU). Her research interests include the modelling and simulation of electro-mechanical interactions in wind farms and multi-objective wind farm control beyond power maximization. Moreover, she is the Centre Manager of FME NorthWind.
\newpage
\begin{wrapfigure}{l}{25mm} 
    \includegraphics[width=0.975in,height=1.3in,clip]{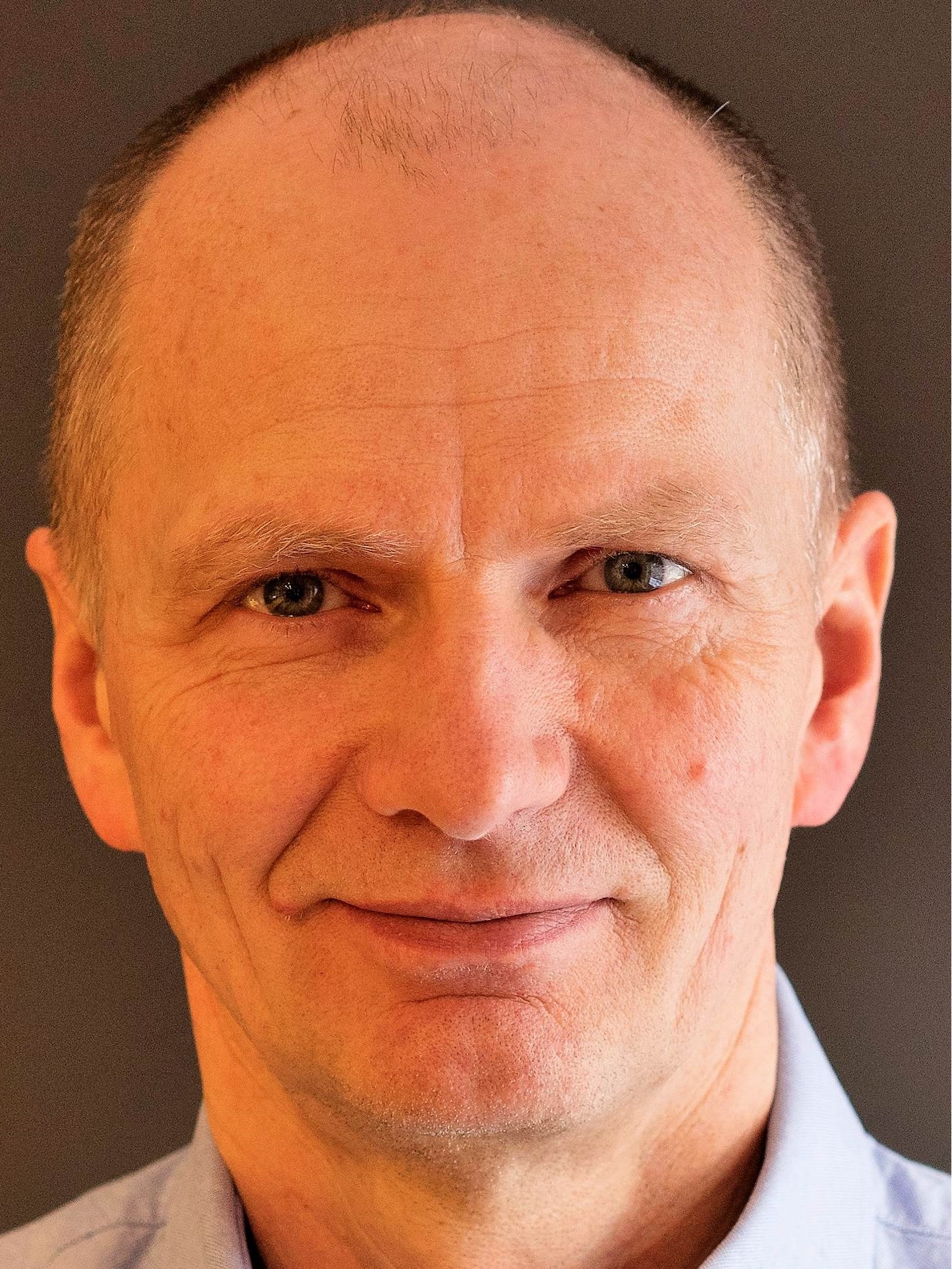}
\end{wrapfigure} \textbf{John Olav Tande} He is Chief Scientist with SINTEF. He has an outstanding track-record of achievements demonstrating excellent capabilities in managing large complex research project in collaboration with industry. He has an extensive international network and strong skills in communication. In May 2019, he was awarded Mission Innovation Champion for his research achievements in offshore wind. Tande was heading FME NOWITECH (2009-2017) and is now heading FME NorthWind (2021-2029)

\begin{wrapfigure}{l}{25mm} 
    \includegraphics[width=0.975in,height=1.3in,clip]{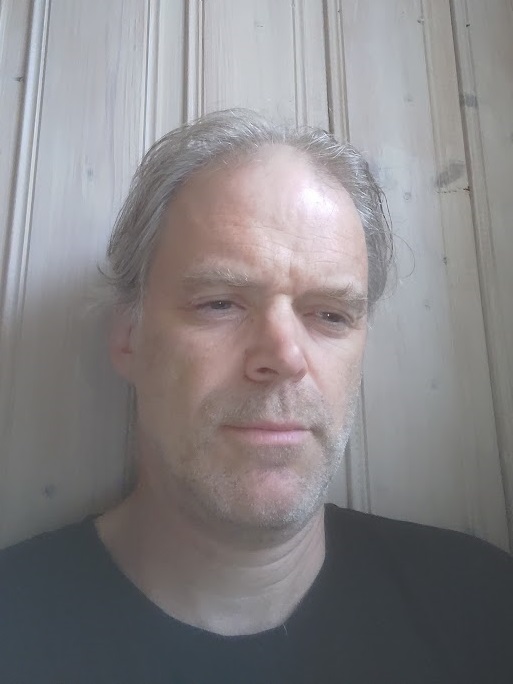}
\end{wrapfigure} \textbf{Idar Barstad} He disciplinary specialist at Kjeller Vindteknikk, Norconsult. He has a Ph.D. in meteorology, 20+ years of experience in academia and 5+ years in consultancy. He has 25 years of experience in using numerical models on global, meso, and local scales. He has led several research initiatives, was involved in the NORCOWE FME (2009-2017) and has guided several students. He contributes in the NorthWind FME through his position in Norconsult.

\begin{wrapfigure}{l}{25mm} 
    \includegraphics[width=0.975in,height=1.3in,clip]{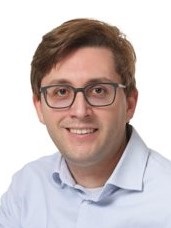}
\end{wrapfigure} \textbf{Alexis Benhamou} He is a Naval Architect in TotalEnergies technical teams dedicated to floating offshore wind development. With its strong background in advanced hydroelastic numerical simulations and structural monitoring of floating structures at Bureau Veritas, he is familiar with the problematics of coupled simulations required for the development of floating offshore wind, in particular to address the digital twin problematics.

\begin{wrapfigure}{l}{25mm} 
    \includegraphics[width=0.975in,height=1.3in,clip]{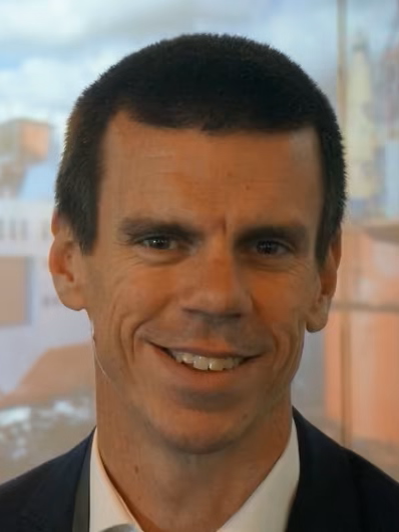}
\end{wrapfigure} \textbf{Thomas Brathaug} He holds an M.Sc. in Naval Architecture from NTNU in Trondheim. He has over 15 years of experience working in ship design and product development for specialized vessels within oil\&gas and renewables. He is currently working in Vard as Principal Naval Architect with responsibility for Vards product portfolio within the renewables segment.

\begin{wrapfigure}{l}{25mm} 
    \includegraphics[width=0.975in,height=1.3in,clip]{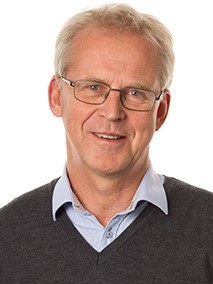}
\end{wrapfigure} \textbf{Tore Christiansen} He has a background in aeronautical engineering and construction management – and thirty years working experience as a researcher, developer, consultant and manager in engineering, information systems and organization science. He has served as designer and project manager of analysis programs and software systems for a range of public and private sector companies in Norway, Europe and the United States. He currently works as Project Coordinator for Floating Wind in the Group Incubator at DNV’s Main Office in Høvik, Norway.

\begin{wrapfigure}{l}{25mm} 
    \includegraphics[width=0.975in,height=1.3in,clip]{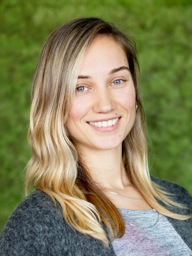}
\end{wrapfigure} \textbf{Anouk-Letizia Firle} She has a Master of Science in Environmental Systems and Resource Management from the University of Osnabrueck in Germany and works as a project and business developer for the government-funded initiative Norwegian Catapult in the company Sustainable Energy. She provides pioneering renewable energy companies with access to test facilities and relevant data from them, such as the Zephyros floating wind turbine, formerly known as the Hywind Demo.

\begin{wrapfigure}{l}{25mm} 
    \includegraphics[width=0.975in,height=1.3in,clip]{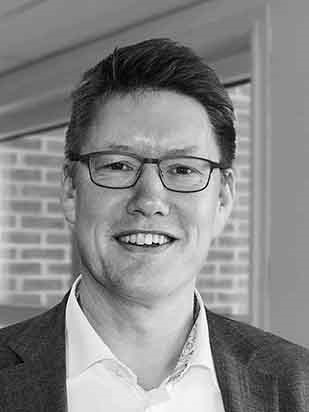}
\end{wrapfigure} \textbf{Alexander Fjeldly} He is the Vice President for Asset Performance Management in FORCE Technology Norway. This involves heading up three departments working on advanced subsea NDT, structural monitoring, and environmental services for reduction of pollution to air. Alexander Fjeldly holds a Dr. Ing. in materials technology from NTNU and has a background and experience from automotive industry, aerospace industry and oil \& gas. He has worked with product development, engineering management and technology leadership for over 25 years.

\begin{wrapfigure}{l}{25mm} 
    \includegraphics[width=0.975in,height=1.3in,clip]{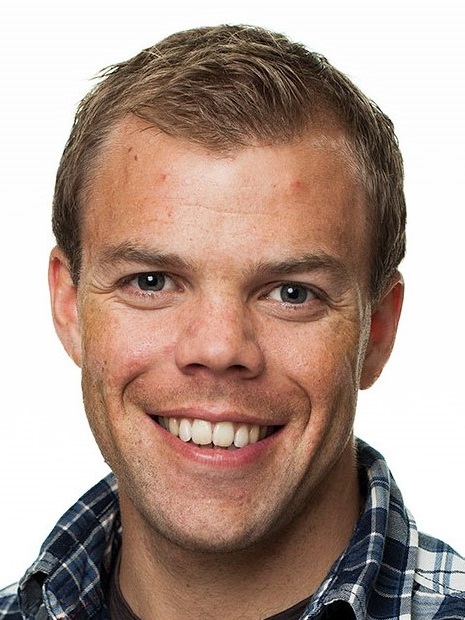}
\end{wrapfigure} \textbf{Lars Frøyd} He has a PhD in offshore wind technology from NTNU in 2012. Since then, he has been working in 4Subsea and is now Lead Engineer in their Wind Energy Team. He has worked mainly with development of methodology and tools for floating wind turbine analysis, and has had a central role in pioneering floating wind development projects Hywind Scotland and Hywind Tampen

\begin{wrapfigure}{l}{25mm} 
    \includegraphics[width=0.975in,height=1.3in,clip]{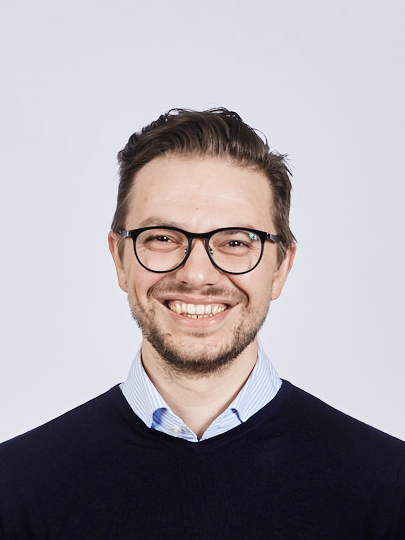}
\end{wrapfigure} \textbf{Alexander Gleim} He has helped customers across heavy-asset industries implement and leverage Cognite Data Fusion and its Digital Twin technology to realize business value at scale. He currently leads Cognite's global Go-To-Market activities. He holds a Master degree in Mathematics from the University of Cambridge and a PhD in Mathematical Statistics from the University of Bonn.
\newpage
\begin{wrapfigure}{l}{25mm} 
    \includegraphics[width=0.965in,height=1.2in,clip]{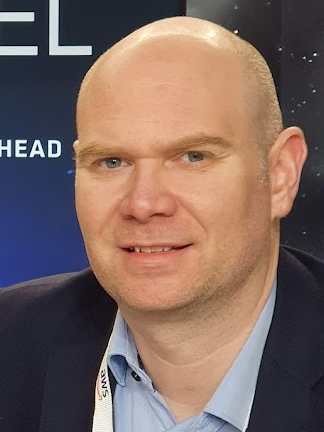}
\end{wrapfigure} \textbf{Alexander Høiberget} He the CBDO at EIDEL, has held various professional roles in technology-driven sectors over the past twenty years. From working in telecom and IT for the enterprise market to electrical engineering for the space and defense sector, he has developed a broad understanding of technology and trends, and how they can address business needs and challenges.

\begin{wrapfigure}{l}{25mm} 
    \includegraphics[width=0.975in,height=1.3in,clip]{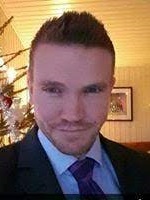}
\end{wrapfigure} \textbf{Håvard Paulshus} He currently serves as the Director of Solutions for Renewables and Utilities at Kongsberg Digital, where he brings a wealth of expertise in Digitalization, Engineering, and Management garnered from years of experience in the Energy Industry. His educational foundation includes a Master of Science degree in Mechanical Engineering and Computer Science from NTNU.

\begin{wrapfigure}{l}{25mm} 
    \includegraphics[width=0.975in,height=1.3in,clip]{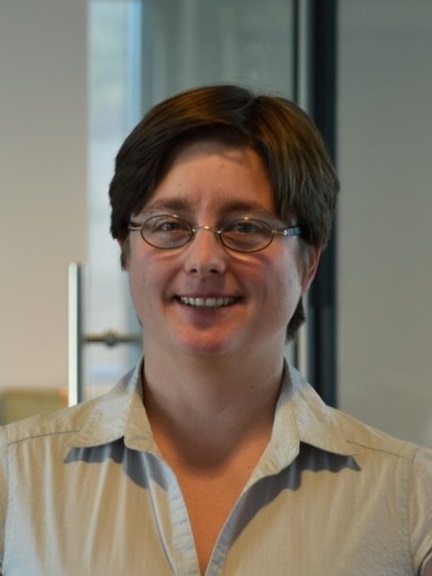}
\end{wrapfigure} \textbf{Catherine Meissner} She holds a PhD in Meteorology from the KIT in Germany. During her career as meteorologist she has been developing mesoscale NWP models and conducted measurement campaigns. Since 2008, her primary working area has been within wind energy. From 2011 onwards she was Software Development Manager at WindSim AS and was responsible for the R\&D activities with the CFD model WindSim. From 2019 to 2021 she worked within energy trading and developed wind and solar models for value. Since 2021 she works for Mainstream Renewable Power as Specialist for Wind Resource assessment focusing on offshore wind farm development.

\begin{wrapfigure}{l}{25mm} 
    \includegraphics[width=0.975in,height=1.3in,clip]{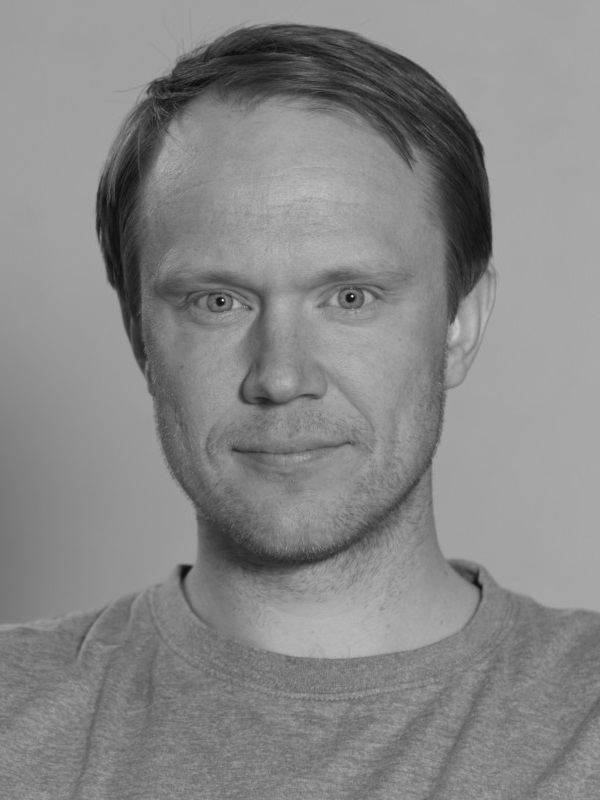}
\end{wrapfigure} \textbf{Guttorm Nygård}, Head of Energy at Store Norske. MSc Financial Economics from NTNU (2014). Experience from finance, business development and strategy with technology and industry companies. Store Norske is a cornerstone company in Longyearbyen with business areas in coal mining, real estate and logistics. Store Norske Energi is a new strategic venture to build a renewable energy company focused on Arctic off-grid settlements.
\newpage
\begin{wrapfigure}{l}{25mm} 
    \includegraphics[width=0.975in,height=1.3in,clip]{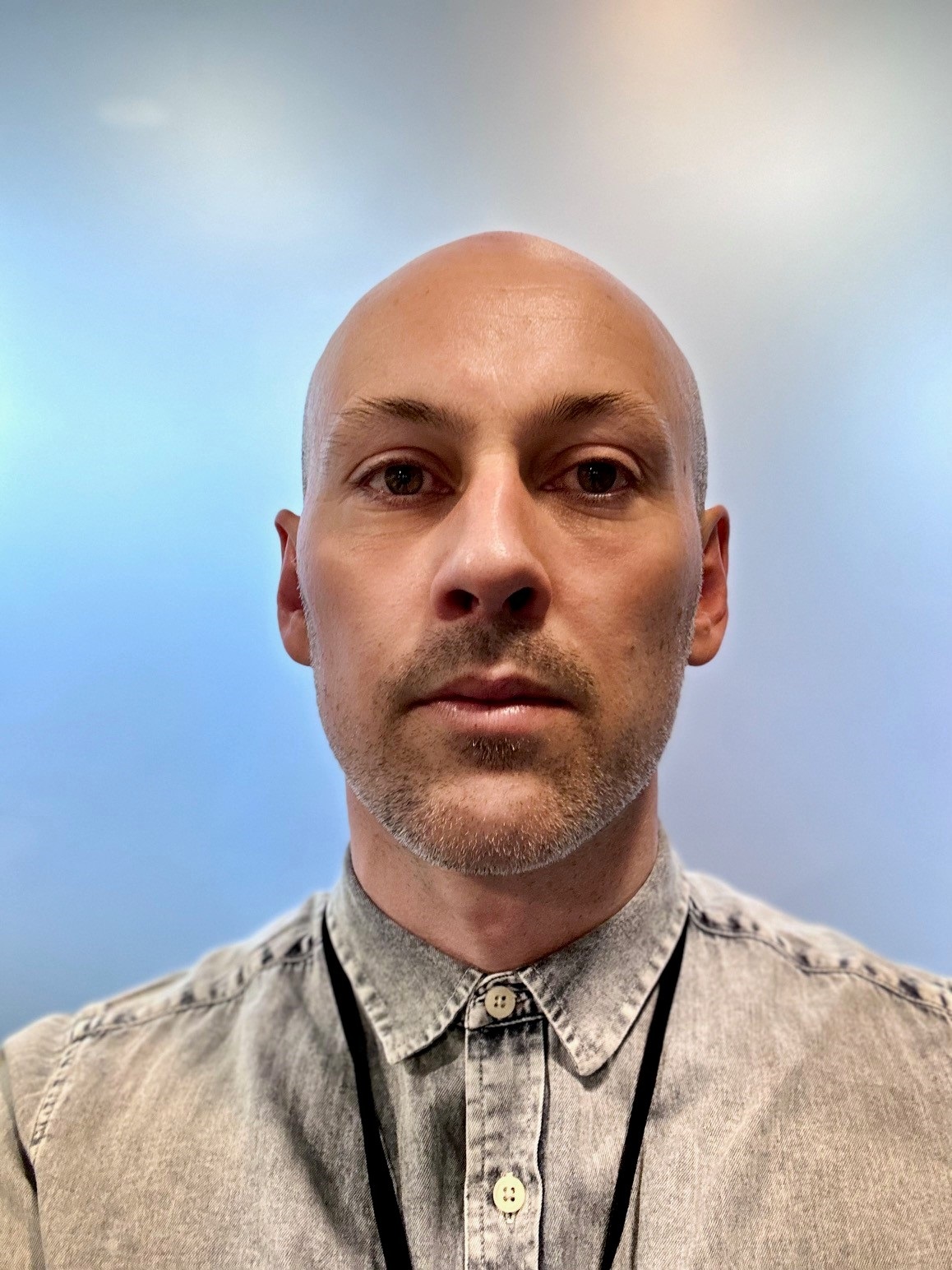}
\end{wrapfigure} \textbf{Jørgen Olsen} is a Senior Systems Engineer at Statkraft. He holds a MSc degree in Energy and Environmental Engineering from NTNU / TU Berlin and has worked with renewable energy since 2008. His position is mainly within data collection, data modelling, system integration and development of user applications for asset performance management.

\begin{wrapfigure}{l}{25mm} 
    \includegraphics[width=0.975in,height=1.3in,clip]{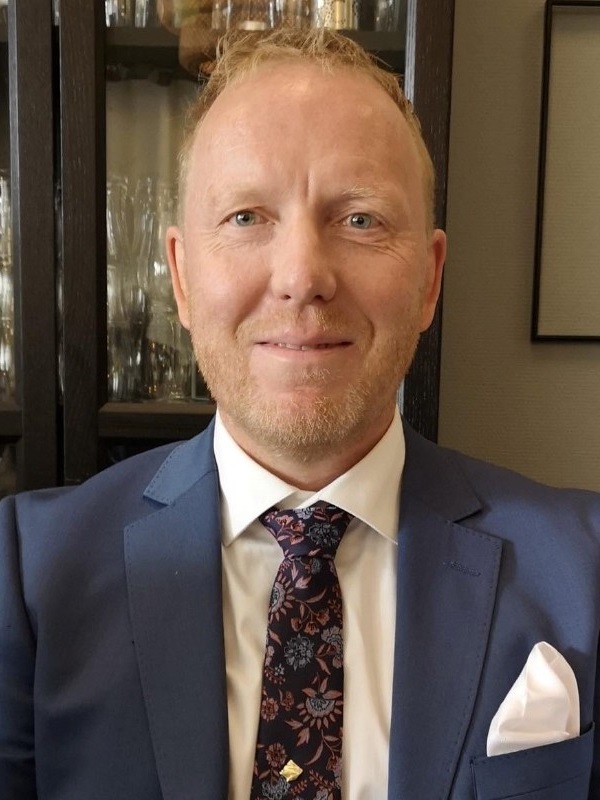}
\end{wrapfigure} \textbf{Tore Rasmussen} In his current position as Principal Engineer-Condition Monitoring and Data Analytics within Aneo, Tore Rasmussen, is responsible to develop condition monitoring systems and data analytics methodologies as part of Aneo’s operation and maintenance strategy. He has been into the wind business since 2008 where he has worked in companies like ScanWind, GE, Kongsberg Digital, Aneo and now Aneo. A selection of his experiences are design, manufacturing and installation of wind turbines, maintenance strategies, prepare for operations projects, technical due diligence, system developments.
Based on his broad experience within the wind business, he knows the value of utilizing operational data in a practical manner to optimize a wind farm owner’s income. For the last 2 years, Tore has been leading a project to develop as system to monitor, analyze and predict failures in wind turbines based on both CMS and SCADA data. Present the system monitors over 200 wind turbines and automatically updates over 50 000 machine learning features daily. The system is already integrated in Aneo’s operational system platform for wind turbines and increases the insight to take predictive decisions to optimize the maintenance activities

\begin{wrapfigure}{l}{25mm} 
    \includegraphics[width=0.975in,height=1.3in,clip]{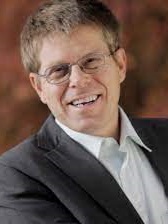}
\end{wrapfigure} \textbf{Elling Rishoff} holds an M.Sc. in Naval Architecture and Ocean Engineering from NTNU (1987).  He has over 35 years’ experience with technology leadership in the marine and technical software fields with a strong know-how in digital transformations. His previous experience includes CEO of DNV Software and DNV Group CIO.  He has engaged with the Offshore wind software industry since 2008. He currently holds the position of Senior Vice President Incubation Offshore Wind at DNV in Norway.”
\newpage
\begin{wrapfigure}{l}{25mm} 
    \includegraphics[width=0.975in,height=1.3in,clip]{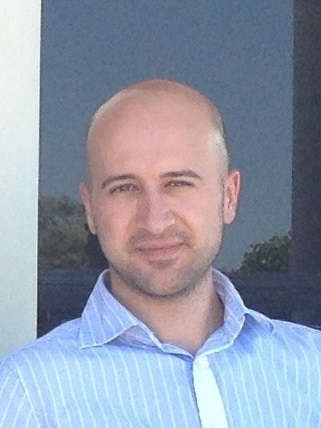}
\end{wrapfigure} \textbf{Francesco Scibilia} holds an M.Sc. in Computer Science Engineering from UNICAL (2006, Italy) and a Ph.D. in Engineering Cybernetics from NTNU (2010, Norway). He is currently technology manager at Equinor, responsible for emerging technologies, foresight and strategy, and innovation within the energy sector. His experience includes academic research and education as PostDoc at the Dept. of Marine Technology at NTNU and as an Adjunct Associate Professor at the Dept. of Engineering Cybernetics at NTNU. He has over 10 years of experience in the industry within technology R\&D and implementation, innovation, project management, and leadership applied to different parts of the energy value chain.

\begin{wrapfigure}{l}{25mm} 
    \includegraphics[width=0.975in,height=1.3in,clip]{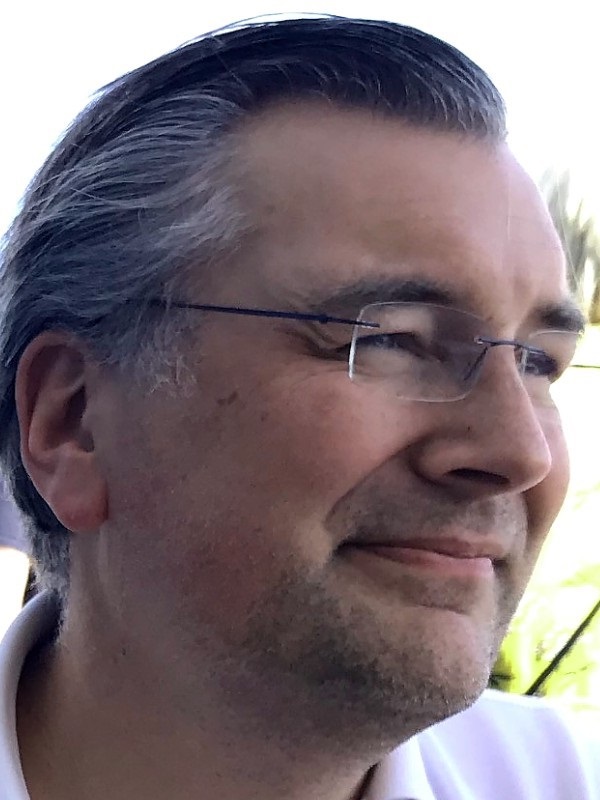}
\end{wrapfigure} \textbf{John Olav Skogås} holds an M.Sc. in Computer Engineering and Telematics from NTNU. He has over 23 years of experience from Norwegian Defence Forces and Kongsberg Maritime AS.  His experience includes working with various forms of data acquisition, processing and analysis, including developing concepts and operational plans. His current position is Lead Engineer within Kongsberg Maritime AS where project management and participation in various research projects within condition monitoring of rotating equipment as well as transition to green fuels take most of the time.

\end{document}